\documentclass[final,3p,times]{elsarticle}

\usepackage{amssymb}
\usepackage{amsmath}
\usepackage{amsthm}
\usepackage{natbib}
\usepackage{amssymb}
\usepackage{mathtools}
\usepackage{multirow}
\usepackage{bm}
\usepackage{stmaryrd}
\usepackage{tabularx}
\usepackage{algorithm}
\usepackage{cancel}
\usepackage{algpseudocode}
\usepackage{tikz}
\usepackage{array}
\usepackage{multirow}
\usepackage{hyperref}

\journal{Elsevier}

\begin{document}

\begin{frontmatter}

\title{\textbf{T}ensor-decomposition-based \textbf{A P}riori \textbf{S}urrogate (TAPS) modeling for ultra large-scale simulations}


\author[a]{Jiachen Guo}
\author[b]{Gino Domel}
\author[b]{Chanwook Park}
\author[a]{Hantao Zhang}
\author[b]{Ozgur Can Gumus}
\author[c]{Ye Lu}
\author[b]{Gregory J. Wagner}
\author[d,e]{Dong Qian}
\author[b]{Jian Cao}
\author[f]{Thomas J.R. Hughes}
\author[b,e]{Wing Kam Liu}

\affiliation[a]{organization={Theoretical and Applied Mechanics Program, Northwestern University},
            addressline={2145 Sheridan Road}, 
            city={Evanston},
            postcode={60201}, 
            state={IL},
            country={USA}}

\affiliation[b]{organization={Department of Mechanical Engineering, Northwestern University},
            addressline={2145 Sheridan Road}, 
            city={Evanston},
            postcode={60201}, 
            state={IL},
            country={USA}}
\affiliation[c]{organization={Department of Mechanical Engineering, University of Maryland, Baltimore County},
            addressline={1000 Hilltop Circle}, 
            city={Baltimore},
            postcode={21250}, 
            state={MD},
            country={USA}}

\affiliation[d]{organization={Department of Mechanical Engineering, University of Texas, Dallas},
            addressline={800 W. Campbell Road}, 
            city={Richardson},
            postcode={75080}, 
            state={TX},
            country={USA}}

\affiliation[e]{organization={Co-Founders of HIDENN-AI, LLC},
            addressline={1801 Maple Ave}, 
            city={Evanston},
            postcode={60201}, 
            state={IL},
            country={USA}}

\affiliation[f]{organization={Institute for Computational Engineering and Sciences, The University of Texas at Austin},
            addressline={201 East 24th Street, Stop C0200}, 
            city={Austin},
            postcode={78712}, 
            state={TX},
            country={USA}}
            
\begin{abstract}
A data-free, predictive scientific AI model, Tensor-decomposition-based A Priori Surrogate (TAPS), is proposed for tackling ultra large-scale engineering simulations with significant speedup, memory savings, and storage gain. TAPS can effectively obtain surrogate models for high-dimensional parametric problems with equivalent zetta-scale ($10^{21}$) degrees of freedom (DoFs). TAPS achieves this by directly obtaining reduced-order models through solving governing equations with multiple independent variables such as spatial coordinates, parameters, and time.   The paper first introduces an AI-enhanced finite element-type interpolation function called convolution hierarchical deep-learning neural network (C-HiDeNN) with tensor decomposition (TD). Subsequently, the generalized space-parameter-time Galerkin weak form and the corresponding matrix form are derived. Through the choice of TAPS hyperparameters, an arbitrary convergence rate can be achieved. To show the capabilities of this framework, TAPS is then used to simulate a large-scale additive manufacturing process as an example and achieves around 1,370x speedup, 14.8x memory savings, and 955x storage gain compared to the finite difference method with $3.46$ billion spatial degrees of freedom (DoFs). As a result, the TAPS framework opens a new avenue for many challenging ultra large-scale engineering problems, such as additive manufacturing and integrated circuit design, among others. 
\end{abstract}




\end{frontmatter}




\begin{table}[h]
\centering
\caption{Nomenclature}
\label{tab:nomenclature}
\begin{tabularx}{\textwidth}{b{2.5cm}|X}
\hline
\textbf{Variables} & \textbf{Description} \\
\hline
\( u^h_e(\bm{x}) \) & Interpolated scalar field defined inside of an element \\ \hline
\( A^{e} \) & Nodes within element $e$ \\ \hline
\( A^{e}_s \) & Nodes within patch domain of element $e$ \\ \hline
$\mathcal{W}_{s,a,p,j}^i(\bm{x}) $ & Convolution patch function at node $j$ for $i$ -th nodal patch with hyperparameters $s$, $a$, and $p$\\ \hline 
\( M \) & Total number of modes  in tensor decomposition (TD) \\ \hline
\( m \) & Index for mode \\ \hline
\( D \) & Total number of dimensions \\ \hline
\( d \) & Index for dimension \\ \hline
\( \bm{x} \) & Independent variable which includes spatial variable \( \bm{x}_s \), parametric variable \( \bm{x}_p \) and temporal variable \( x_t \) \\ \hline
\multirow{2}{2.5cm}{\( \widetilde{N}_d(x_d; a_d, s_d, p_d) \)} & Global C-HiDeNN shape function for dimension \( d \) with dilation parameter \( a_d \), patch size \( s_d \) and reproducing polynomial order \( p_d \) \\ \hline
\( b \) & Source function in laser powder bed fusion process\\ \hline
\( u^{TD} \) & Approximation of the solution field expressed via TD \\ \hline
\( \mathcal{T} \) & Time slab index for space-parameter-time problem  \\ \hline
\( k \) & Thermal conductivity \\ \hline
\( \rho \) & Material density \\ \hline
\( c_p \) & Heat capacity \\ \hline
\( \eta \) & Material absorptivity \\ \hline
\( P \) & Laser power \\ \hline
\( r \) & Standard deviation that characterizes the width of the heat source\\ \hline
\( q \) & Heat flux\\ \hline
\( q_{conv} \) & Heat flux from convection\\ \hline
\( q_{rad} \) & Heat flux from radiation\\ \hline
\( q_{evap} \) & Heat flux from evaporative cooling\\ \hline
\(h_{conv}\) & Convection coefficient\\ \hline
\(\sigma_{SB}\) & Stefan-Boltzman constant\\ \hline
\(m_{evap}\) & Mass evaporation flux\\ \hline
\(L_{evap}\) & Heat of evaporation\\ \hline
\(\widetilde{B}_{n_d}\) & Shape function derivative \\ \hline
\( \mathbb{U}_d \) & Solution matrix (\( \mathbb{R}^{n_d \times M} \)) for dimension \( d \) that contains all the modes \\ \hline
\end{tabularx}
\end{table}

\section{Introduction}
\label{sec1}

Precision is a fundamental aspect of scientific and engineering applications, especially in advanced industries such as semiconductor manufacturing. The capability to perform accurate computational simulations for these applications is essential for advancing these fields. Precise simulations enable the optimization of design and manufacturing processes by utilizing virtual prototypes and process simulations. This reduces the need for expensive physical prototypes and tests and provides virtual prototypes in circumstances where physical ones are impractical. Traditional computational methods for engineering simulations, however, suffer from prohibitive computational costs when attempting to accurately predict responses across multiple length and time scales (typically done by increasing mesh resolution), making achieving high precision for large-scale problems challenging. In fact, the random-access memory (RAM) requirement can be far beyond the capability of typical workstations and may require massive parallelization on supercomputers. In other industries, such as additive manufacturing (a term encompassing all forms of 3D printing), the vast design space further exacerbates these limitations, as numerous expensive simulations are required to thoroughly explore the effects of different design parameters. 

To fulfill the ever-growing challenges in predictive scientific models, data-driven surrogates, especially artificial intelligence (AI)-based models, present an alternative to conventional numerical models by significantly reducing the forward prediction time. These models can be treated as a reasonably accurate, reduced representation of real physics. Once trained properly, they can be used for fast prediction on unseen parameters \cite{li2020fourier, huang2023introduction}.  However, it is still uncertain whether a data-driven surrogate model can be trained to achieve the level of accuracy required in engineering design. Recently, it has been pointed out by Wolfram Research that standard AI models cannot easily fulfill the high accuracy requirement of predictive scientific tasks \cite{Wolfram_2024}. Furthermore, as suggested by Google Deepmind, the real potential of AI models lies in enhancing, rather than thoroughly replacing, well-established classical numerical algorithms \cite{deepmindGoldenDiscovery}. In addition, the current standard data-driven approaches follow an offline-online scheme, where the offline stage involves a huge amount of training data, which can again be prohibitive. For problems with known physics, this data can be obtained by running multiple expensive simulations relying on standard numerical algorithms. In scenarios involving high-dimensional design spaces governed by parameterized partial differential equations (PDEs), such as in additive manufacturing (AM), conducting repetitive simulations with varying parameters in this offline stage becomes exceedingly expensive both in terms of computation time and data storage.


\begin{figure}[!hbt]
\centering
\includegraphics[width=0.8\linewidth]{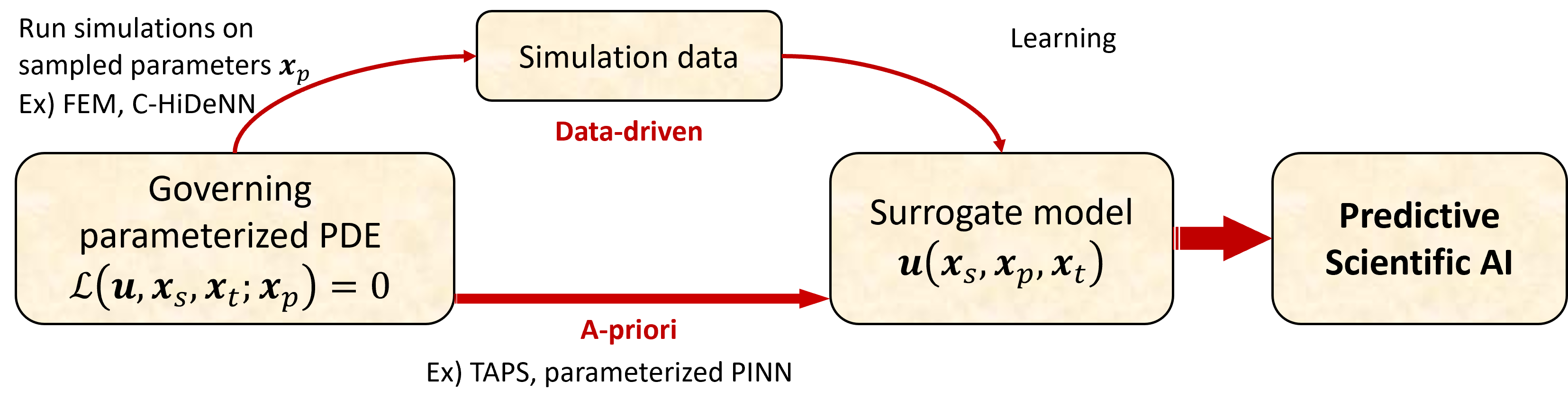}
\caption{The parameterized PDE is a PDE that includes parameters $\bm{x}_p$ that can vary and influence the solution $\bm{u}(\bm{x}_s,\bm{x}_p,{x}_t)$, where $\bm{x}_s$ and ${x}_t$ are the spatial and time variables, respectively. The a priori approach directly finds a surrogate model from the governing parameterized PDE, whereas the data-driven approach has to solve the parameter-fixed PDE on sampled parameters to generate simulation data, followed by training tasks. FEM: Finite Element Method \cite{liu2022eighty}, C-HiDeNN: Convolution Hierarchical Deep-learning Neural Network \cite{lu2023convolution}, TAPS: Tensor-decomposition-based A Priori Surrogate, PINN: Physics Informed Neural Network \cite{raissi2019physics}.}
\label{AP}
\end{figure}
To avoid the prohibitive offline stage, one can try to obtain a surrogate model directly from governing equations without generating any data. As shown in Fig. \ref{AP} denoted by the words ``A Priori", this approach aims to find the surrogate model before actually ``seeing" any data. 
For example, multilayer perceptron (MLP) architectures have been vastly used in physics-informed neural networks (PINNs) and their variations to approximate solutions to PDEs without requiring data \cite{raissi2019physics, zhang2022analyses}. However, the results of these efforts are underwhelming, as it has been shown that PINN results have often been compared to weak baselines \cite{mcgreivy2024weak}, and it is unclear if they guarantee convergence. Moreover, this method is still susceptible to high computational costs for both large-scale and high-dimensional problems \cite{cho2024separable}. 

\begin{figure}[!hbt]
\centering
\includegraphics[width=1.0\linewidth]{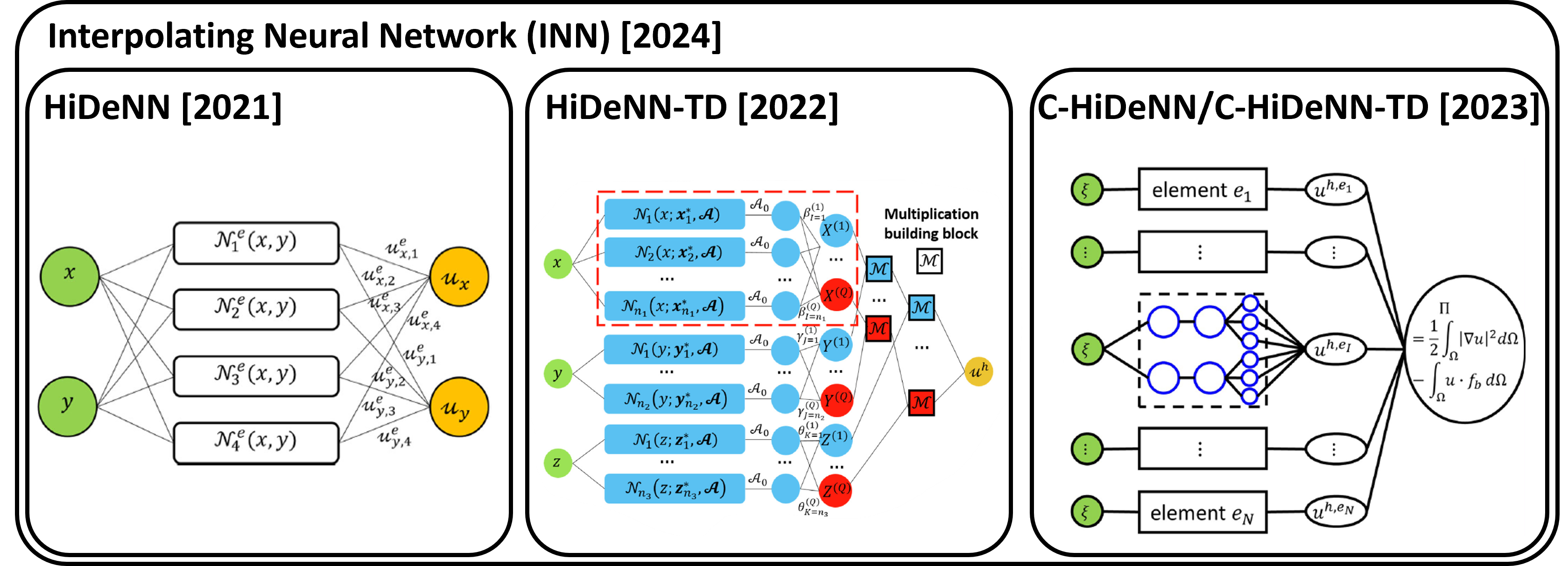}
\caption{Development history of INN \cite{park2024engineering}. Figures are borrowed from references: HiDeNN \cite{zhang2021hierarchical}, HiDeNN-TD \cite{zhang2022hidenn}, C-HiDeNN \cite{park2023convolution}, C-HiDeNN-TD \cite{lu2023convolution}.}
\label{fig:INN_history}
\end{figure}

Instead of developing solvers solely based on machine learning concepts, a new class of \textbf{Hi}erarchical \textbf{De}ep-learning \textbf{N}eural \textbf{N}etworks (HiDeNN) has been developed recently. This network architecture incorporates principles from the finite element method (FEM) to construct their architecture \cite{saha2021hierarchical, zhang2021hierarchical}. Originally designed to advance FEM as opposed to solve parameterized PDEs, this approach significantly enhances computational accuracy and efficiency for both linear and nonlinear problems compared to standard FEM \cite{liu2023hidenn}. HiDeNN was then enhanced by adding an additional hidden layer in the form of a nonlinear convolutional filter, formulating a new neural network architecture named \textbf{C}onvolutional HiDeNN (C-HiDeNN) \cite{lu2023convolution, park2023convolution}. C-HiDeNN mimics the structure of the generalized finite element method but leverages machine learning to optimize its hyperparameters to further improve accuracy and efficiency. Arbitrary orders of convergence have been observed for C-HiDeNN despite utilizing a linear finite element mesh \cite{park2023convolution}. Although these methods offers greater accuracy with fewer DoFs, like FEM, they still encounter computational challenges such as balancing memory usage against mesh resolution, which limits their efficiency in modeling ultra large-scale and high-dimensional problems. Therefore, it becomes necessary to employ model order reduction techniques to address these limitations.

Model order reduction techniques have been widely used to tackle the ever-growing challenges from high-dimensional and large-scale problems. For example, proper generalized decomposition (PGD) \cite{chinesta2011overview, chinesta2013proper, nouy2010priori} has been proposed to efficiently solve high-dimensional PDEs. Recently, tensor decomposition (TD) has been successfully leveraged within the HiDeNN framework. For example, Zhang showed that HiDeNN combined with TD (HiDeNN-TD) significantly improved the speed of HiDeNN while maintaining higher accuracy \cite{zhang2022hidenn}. Li proposed C-HiDeNN combined with TD (C-HiDeNN-TD) for extremely large-scale nested topology optimization problems \cite{li2023convolution}. Recently, Park generalized the HiDeNN-family networks under the umbrella of Interpolating Neural Networks (INNs) and demonstrated that the network can be used for both data-driven learning and data-free (i.e., a priori) solving \cite{park2024engineering}. The development history of HiDeNN family networks and INN is summarized in Fig. \ref{fig:INN_history}. While INN clearly explains how to construct the network architecture, an efficient optimization scheme for solving ultra large-scale and high-dimensional problems remains underdeveloped. In this paper, ultra large-scale problems refer to problems on the zetta-scale ($10^{21}$) in terms of DoFs.

The demand for high-precision engineering simulations and efficient solution schemes highlights the need for innovative modeling approaches that swiftly solve large-scale problems while optimizing the design space. This research aims to fulfill this need by developing tensor-decomposition-based A Priori Surrogate (TAPS), a data-free predictive AI model, which aims to enhance high-resolution capabilities while simultaneously optimizing computational efficiency with a minimal memory footprint, low data storage needs, and fast prediction. The proposed comprehensive framework sets a foundation for scalable, adaptable, and future-proof solutions to counter the ever-growing complexity in simulation-driven advanced industries. TAPS is particularly well-suited for engineering challenges where: 1) the finite element method and other conventional methods are unsuitable due to excessively long simulation times or high RAM and storage demands needed to achieve high accuracy, 2)  the model must accommodate design parameters as inputs, or 3) fast prediction is required once the model is obtained.

This paper is structured as follows. We first introduce the formulation of TAPS in section 2. In section 3, we examine the numerical convergence of TAPS for both space-time (S-T) and space-parameter-time (S-P-T) problems (i.e., problems that are dependent on spatial, parametric, and temporal inputs). In section 4, TAPS is applied to large-scale additive manufacturing problems that are considered intractable with standard numerical algorithms. This application effectively demonstrates TAPS's capability to address all of the three identified challenges.

\section{Theory}
\subsection{Review of C-HiDeNN interpolation theory}
\label{chidenn}

Leveraging the universal approximation theorem, multilayer perceptrons (MLPs) have been successfully applied as global basis functions in deep learning-based solvers \cite{raissi2019physics}. However, as shown in Table \ref{tab:mlp_chidenn}, MLPs have a few potential caveats when approximating PDE solutions. To overcome these limitations, we leverage the Convolutional HiDeNN (C-HiDeNN) interpolation function, which leverages the merits of both locally supported finite element shape functions and the flexibility of machine learning. Note that C-HiDeNN also belongs to the INN category as shown in Fig. \ref{fig:INN_history}. C-HiDeNN maintains all the essential finite element approximation properties such as  Kronecker delta and partition of unity \cite{park2023convolution}.  

\begin{table}[hbt!]
\small
\centering

\caption{Comparison of MLP and C-HiDeNN as approximation functions of PDE solutions.}
\label{tab:mlp_chidenn}
\begin{tabular}{l|l|l}

\hline
 & \textbf{MLP} & \textbf{C-HiDeNN} \\ \hline
\textbf{Boundary/initial condition} & Penalty term in the loss function \cite{raissi2019physics} & Automatic satisfaction \cite{lu2023convolution} \\ 
\textbf{Convergence and stability} & Stochastic in nature and not guaranteed & Shown for different PDEs \cite{lu2023convolution} \\ 
\textbf{Numerical integration} & Quasi-Monte Carlo integration \cite{kharazmi2021hp} & Gaussian integration \cite{hughes2003finite} \\ 
\textbf{Interpretability} & Black-box model & Interpretable \cite{park2024engineering} \\ \hline
\end{tabular}

\end{table}

\begin{figure}[!hbt]
\centering
\includegraphics[width=0.8\linewidth]{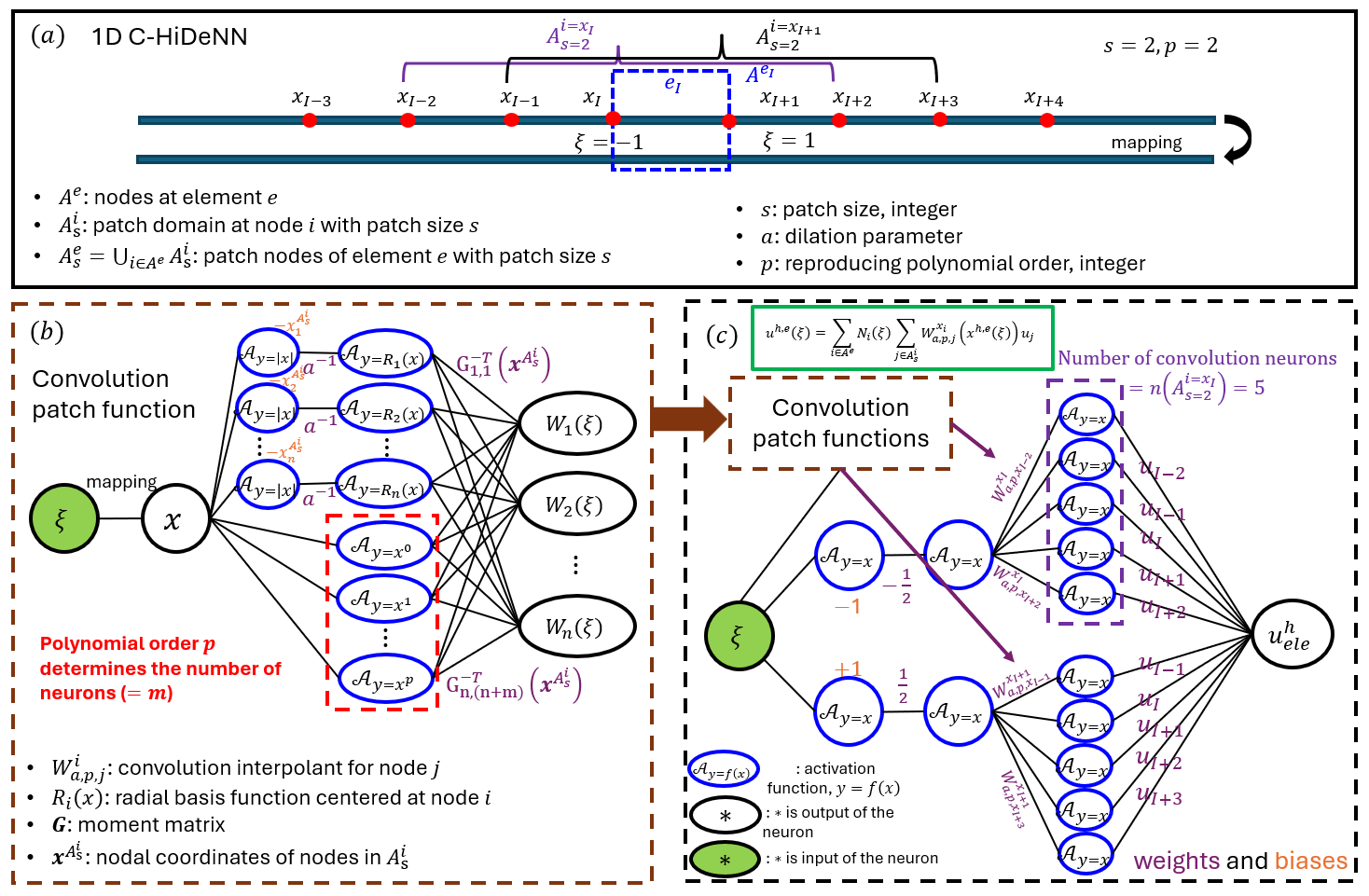}
\caption{(a) Covolution patch in 1D C-HiDeNN shape function (b) Construction of convolution patch function (c) C-HiDeNN shape function as MLP with 3 hidden layers}
\label{shape_fun}
\end{figure}

We first review the C-HiDeNN formulation as illustrated in Fig. \ref{shape_fun} (a) \cite{park2023convolution}. A scalar field $u(\bm{x})$ defined in each element within a domain $\Omega_{\bm{x}}$ can be approximated using C-HiDeNN interpolation as:
\begin{equation}
    u_{e}^{h}(\bm{x})=\sum_{i\in A^{e}}N_{i}(\bm{x})\sum_{j\in A_{s}^{i}}\mathcal{W}_{s,a,p,j}^{i}(\bm{x})u_{j}=\sum_{k\in A_{s}^{e}}\widetilde{N}_{k}(\bm{x};s,a,p)u_{k}
\end{equation}
where $u_j$  is the nodal value and $u_j=u(\bm{x}_j)$; $N_i$ is the linear finite element shape function at node $j$ centered in $i$-th nodal patch; $\mathcal{W}_{s,a,p,j}^i$ is the convolution patch function at node $i$ that can be represented with a partially connected MLP as illustrated in Fig. \ref{shape_fun} (b). The convolution patch functions are controlled by three hyperparameters: patch size $s$ that controls nodal connectivity, dilation parameter $a$ that normalizes distances between patch nodes, and reproducing order $p$ that defines types/orders of activation functions to be reproduced by the patch functions. Due to the inherent local support nature of both $N_i$ and $\mathcal{W}_{s,a,p,j}^i$, the C-HiDeNN shape function  $\widetilde{N}_{k}(\bm{x};s,a,p)$ is also locally supported.

Similar to standard finite element, the approximation for the solution field can be written as:
\begin{equation}
    u^{h}(\bm{x})=\sum_{k} ^ {nnode}\widetilde{N}_{k}(\bm{x};s_{k},a_{k},p_{k})u_{k}
\label{full_chidenn}
\end{equation}
where $nnode$ is the total number of nodes and $k$ is the nodal index.
It should be noted that the hyperparameters $s,a,p$ can vary across nodes since C-HiDeNN can optimize these hyperparameters like machine learning parameters, rendering an adaptable functional space without altering the number of global nodes or hidden layers. This clearly distinguishes C-HiDeNN from MLP, where the activation functions and network architectures are mostly fixed.

The C-HiDeNN shape function $\widetilde{{N}}_k(\bm{x})$ satisfies Kronecker-delta property at nodal positions \cite{lu2023convolution} (hyperparameters $s,a,p$ are dropped for brevity):
\begin{equation}
    \widetilde{N}_{k}(\bm{x}_{l})=\delta_{kl}
\end{equation}
where the Kronecker delta is defined as:
\begin{equation}
    \delta_{kl}=\left\{\begin{array}{cc}0& \quad \mathrm{if} \ k\neq l,\\1& \quad \mathrm{if} \ k=l.\end{array}\right.
\end{equation}
Thus, at the Dirichlet boundary node $\bm{x}_b$ where $u(\bm{x}_b) = u_b$, C-HiDeNN automatically satisfies the Dirichlet boundary condition:
\begin{equation}
    u^{h}(\bm{x}_b) = \sum_{k}^{nnode}\widetilde{N}_k(\bm{x}_b)u_k=u_b
\end{equation}
Going forward, we will employ the C-HiDeNN shape function $\widetilde{N}_{k}(\bm{x})$ as the locally supported basis function for the interpolation.

\subsection{Discrete Tensor decomposition}
Tensor decomposition is a mathematical technique used to break down a high-dimensional tensor, such as a 3D finite element solution, into a set of simpler components, making it easier to analyze, store, and process \cite{kolda2009tensor}. It generalizes matrix decomposition methods like singular value decomposition (SVD) to higher-order tensors.

\begin{figure}[!hbt]
\centering
\includegraphics[width=0.8\linewidth]{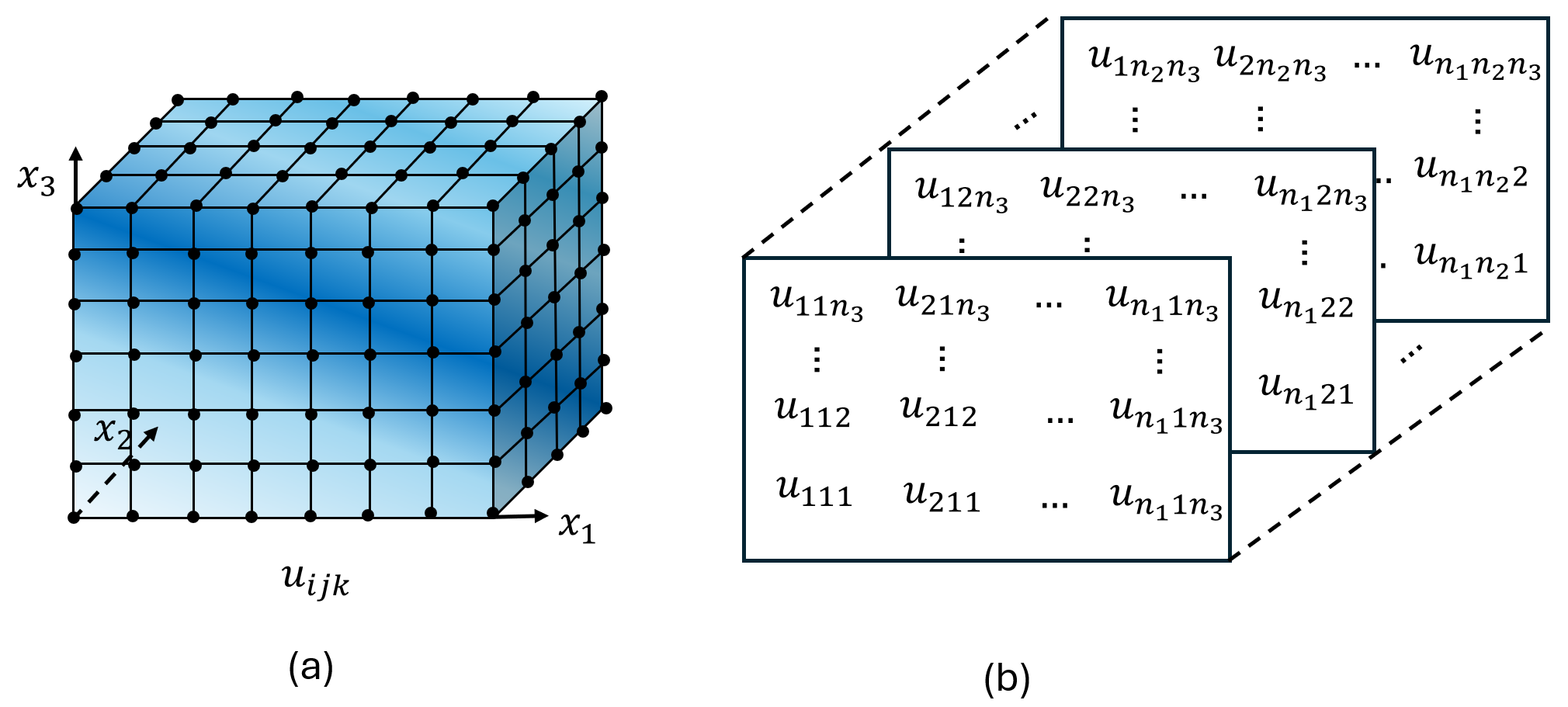}
\caption{(a) 3D Cartesian mesh. (b) Nodal values can be treated as a 3rd order tensor.}
\label{fig: discrete_TD}
\end{figure}

Consider a cubic spatial domain $\Omega_{\bm{x}}$ discretized with a regular Cartesian grid where each grid point (or node) stores a scalar value (see Fig. \ref{fig: discrete_TD}). The discrete nodal values can be represented as a 3rd order tensor $u_{IJK}$ where $I=1,..,n_1;J=1,…,n_2;K=1,…,n_3$. The number of DoFs for this structured mesh is $n_1\times n_2\times n_3$. When high resolution is required for the analysis, as the case in AM simulations, the number of DoFs can be extremely large. To effectively reduce the DoFs, different discrete tensor decomposition methods can be used to project the original 3rd order tensor into lower order tensors. In this paper, we focus on CANDECOMP/PARAFAC (CP) decomposition, where the higher-order tensors are approximated using a finite sum of products of 1D vectors \cite{kolda2009tensor}:
\begin{equation}
    u_{IJK}\approx u_{IJK}^{TD}=\sum_{m=1}^{M}u_{Im}^{[1]}u_{Jm}^{[2]}u_{Km}^{[3]}
\label{3d_td}
\end{equation}
where $M$ is defined as the total number of modes in CP decomposition; $u_{I m}^{[1]}$ refers to the projected 1D vector in the first dimension and $m$-th mode; the superscript $[d]$ represents the dimension index and $d=1,2,3$; the 1st subscript $I$ is the nodal index, and the 2nd subscript $m$ refers to the modal index.

\begin{figure}[!hbt]
\centering
\includegraphics[width=0.8\linewidth]{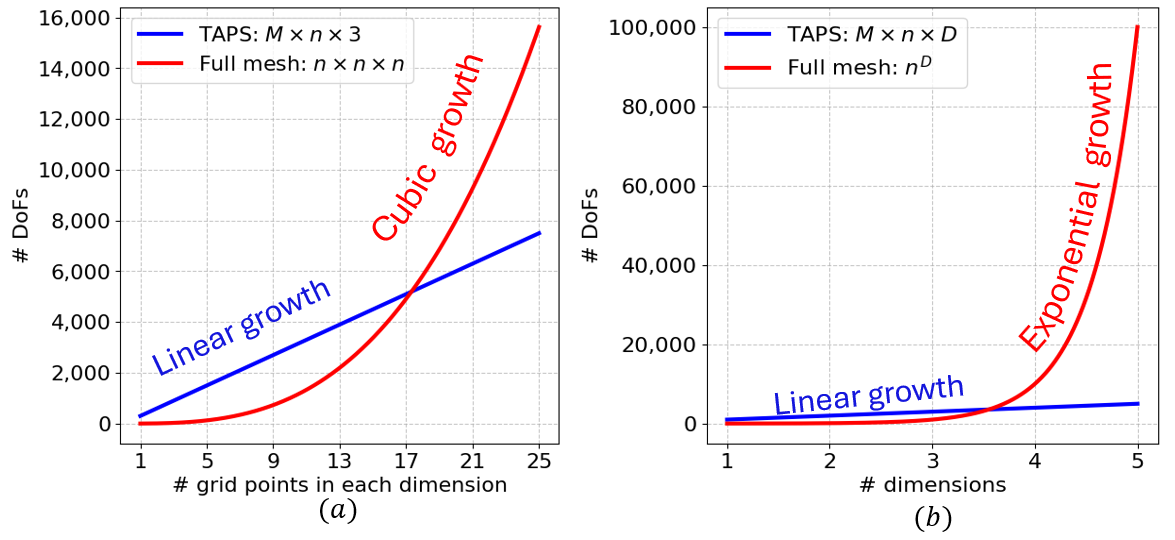}
\caption{Comparison of number of DoFs, (a) in terms of mesh size $n$, (b) in terms of problem dimension $D$}
\label{growth}
\end{figure}

As can be seen from Eq. \ref{3d_td}, with CP decomposition, the total number of DoFs can be reduced from $n_1\times n_2\times n_3$ to $M\times (n_1+n_2+n_3)$. Assuming $M$ does not increase when the mesh is refined along each dimension, then the solution matrix $u_{IJK}$ will have cubic growth, whereas CP decomposition $\sum_{m=1}^{M}u_{Im}^{[1]}u_{Jm}^{[2]}u_{Km}^{[3]}$ only exhibits linear growth, as shown in Fig. \ref{growth} (a). This reduction is paramount to making large-scale simulation achievable.

As an extension of the previous case, we consider $D$ dimensional general time-dependent parametric problems where the independent variables $(x_1,x_2,…,x_D)$ can be classified into 3 different categories, namely, spatial variables $\bm{x}_s$, parametric variables $\bm{x}_p$, and temporal variable $x_t$. Spatial variables $\bm{x}_s$ describe the spatial coordinates of the problem. Parametric variables $\bm{x}_p$ can represent any PDE coefficients, initial/boundary conditions, or geometry descriptors as extra-coordinates. The temporal variable $x_t$ represents time. Assuming the spatial domain $\Omega_{\bm{x}_s}$ is cubic, the parametric domain $\Omega_{\bm{x}_p}$ is hypercubic and Cartesian grids are used for discretization, then the nodal solution to these problems can be written as a discrete $D$-th order tensor $u_{I_{1}I_{2},...,I_{D}}$. Similarly, CP decomposition can be used to effectively decompose higher-order tensors into a finite sum of tensor products of 1D vectors.

\begin{equation}
    u_{I_{1}I_{2},...,I_{D}}\approx u_{I_{1}I_{2},...,I_{D}}^{TD}=\sum_{m=1}^{M}u_{I_{1}m}^{[1]}u_{I_{2}m}^{[2]}...u_{I_{D}m}^{[D]}
\label{nd_td}
\end{equation}
If every dimension is discretized into $n$ grid points, then a $D$-th order tensor will have DoFs of $n^D$, whereas CP decomposition only requires $M\times D\times n$ DoFs. Consequently, CP decomposition can dramatically reduce the total DoFs of general high-dimensional parametric problems, as shown in Fig. \ref{growth} (b). 

\subsection{TD interpolation in TAPS}

Assume that the $D$-th order tensor $u_{I_{1}I_{2},...,I_{D}}$ represents a $D$-input one-output continuous function $u(\bm{x})$ measured at a Cartesian grid discretized with $I_1,I_2,...,I_D$ grid points in each input dimension. The discrete tensor decomposition $u_{I_{1}I_{2},...,I_{D}}^{TD}$ can only approximate the function $u(\bm{x})$ at these grid points. In this case, how can we measure the value of the function on an arbitrary input $\bm{x}$ with tensor decomposition? A natural answer is using C-HiDeNN interpolation functions.

Similar to standard finite element shape functions, for a 3D spatial problem discretized with a Cartesian grid, a 3D C-HiDeNN interpolation function can be rewritten as a tensor product of one-dimensional C-HiDeNN interpolation functions (hyperparameters $s,a$ and $p$ will be dropped from now on for brevity):
\begin{equation}
    \widetilde{N}_{k}(x_1, x_2, x_3)=\widetilde{N}_{I}^{[1]}(x_1)\widetilde{N}^{[2]}_{J}(x_2)\widetilde{N}^{[3]}_{K}(x_3)
\end{equation}
where the superscript refers to the dimension of the 1D C-HiDeNN shape function. Therefore, we can rewrite Eq. \ref{full_chidenn}  as:

\begin{equation}
    u^{h}(\bm{x_s})=\sum_{I} \sum_{J} \sum_{K} \widetilde{N}^{[1]}_{I}(x_1)\widetilde{N}^{[2]}_{J}(x_2)\widetilde{N}^{[3]}_{K}(x_3)u_{IJK}
\label{3d_chidenn}
\end{equation}
where $\bm{x}_s = \left[ x_1,x_2,x_3\right] $ is the spatial variable. Plugging the CP decomposition form of the tensor $u_{IJK}^{TD}$ into Eq. \ref{3d_td} into Eq. \ref{3d_chidenn} and rearranging the terms, we have:

\begin{equation}
    u^{TD}(\bm{x_s})=\sum_{m=1}^M \left[\sum_{I}  \widetilde{N}^{[1]}_{I}(x_1)u_{Im}^{[1]}\right]\left[\sum_{J}\widetilde{N}^{[2]}_{J}(x_2)u_{Jm}^{[2]}\right]\left[\sum_{K} \widetilde{N}^{[3]}_{K}(x_3)u_{Km}^{[3]}\right]
\label{td_3d}
\end{equation}
Eq. \ref{td_3d} represents the TD interpolation (with C-HiDeNN) for a 3D spatial problem. Extending this framework to a general $D$-dimensional space-parameter-time (S-P-T) problem with independent variables defined in Eq. \ref{variables}:

\begin{equation}
    \bm{x}=(\underbrace{x_1,...,x_{S}}_\textrm{spatial variables}, \underbrace{x_{S+1},...,x_{P}}_\textrm{parametric variables}, x_t)
\label{variables}
\end{equation}
Then the TD interpolation to the S-P-T solution field can be written as follows:
\small
\begin{equation}
    u^{TD}(\bm{x}_s, \bm{x}_p,{x}_t)=\sum_{m=1}^M \underbrace{\left[\sum_{I_1}  \widetilde{N}^{[1]}_{I_1}(x_1)u_{I_1m}^{[1]}\right]\cdots\left[\sum_{I_S}\widetilde{N}^{[S]}_{I_S}(x_{I_S})u_{I_Sm}^{[S]}\right]}_\textrm{spatial }\underbrace{\left[\sum_{I_{S+1}} \widetilde{N}^{[S+1]}_{I_{S+1}}(x_{S+1})u_{I_{S+1}m}^{[S+1]}\right]\cdots \left[\sum_{P} \widetilde{N}^{[P]}_{I_P}(x_P)u_{I_Pm}^{[P]}\right]}_\textrm{parametric} \underbrace{\left[\sum_{I_D} \widetilde{N}^{[D]}_{I_D}(t)u_{I_D}^{[D]}\right]}_\textrm{temporal}
\label{td_eq_0}
\end{equation}
\normalsize
This can be further simplified using the product notation:
\begin{equation}
    u^{TD}(\bm{x}_s, \bm{x}_p,{x}_t)=\sum_{m=1}^M\prod_{d=1}^D\sum_{I_d}\widetilde{N}_{I_d}^{[d]}(x_d)u_{I_dm}^{[d]}
\label{td_eq}
\end{equation}
where $\widetilde{N}_{I_d}^{[d]}(x_d)$ refers to the 1D C-HiDeNN shape function in the $d$-th dimension; $u_{I_dm}^{[d]}$ is the nodal solution for dimension $d$ and mode $m$.

\subsection{The General S-P-T Galerkin form of TAPS}

Similar to FEM, TAPS adopts the weighted-sum formulation to solve PDEs.  Consider a general S-P-T PDE:

\begin{equation}
\label{eq:pde}    \mathcal{L}\left(u\left(\bm{x}\right)\right)=f(\bm{x}),
\end{equation}
where $\mathcal{L}$ is the differential operator; the independent variable vector $\bm{x} = ({\bm{x}_s, \bm{x}_p, x_t})$; $f(\bm{x})$ is the forcing function. Table \ref{tab:pde} lists different examples of operator $\mathcal{L}$ and corresponding dependent and independent variables.

\begin{table}[hbt!]
\small
\caption{Examples for differential operators, dependent and independent variables}
\label{tab:pde}
\centering{%
\begin{tabular}{p{5cm}|c|c|c|c|c}
\hline
PDE & Differential operator $\mathcal{L}$ & Dependent variable &$\bm{x}_s$ & $\bm{x}_p$ & $x_t$ \\ \hline
$\frac{\partial^{2}u}{\partial x_{1}^{2}}+\frac{\partial^{2}u}{\partial x_{2}^{2}}+...+\frac{\partial^{2}u}{\partial x_{D}^{2}} = f(\bm{x})$   &   $\frac{\partial^{2}}{\partial x_{1}^{2}}+\frac{\partial^{2}}{\partial x_{2}^{2}}+...+\frac{\partial^{2}}{\partial x_{D}^{2}}$   & $u$    & $(x_1,x_2,...,x_D)$  &    -        &   -    \\ \hline
 $\mu u_{i,jj}+(\mu+\lambda)u_{j,ij}+F_i=e^{x_1^2 + x_2^2 + x_3^2}$    &    $\mu (\cdot)_{i,jj}+(\mu+\lambda)(\cdot)_{j,ij}$ & $u_i, \quad i=1,2,3$    &  $(x_1, x_2, x_3) $    &   $(\lambda, \mu)$        &  -  \\   \hline
 
 $\rho c_p\frac{\partial u}{\partial t} + k(\frac{\partial^{2}u}{\partial x_{1}^{2}}+\frac{\partial^{2}u}{\partial x_{2}^{2}}+\frac{\partial^{2}u}{\partial x_{3}^{2}}) = P e^{x_1^2 + x_2^2 + x_3^2}$ & $\rho c_p\frac{\partial}{\partial t} + k(\frac{\partial^{2}}{\partial x_{1}^{2}}+\frac{\partial^{2}}{\partial x_{2}^{2}}+\frac{\partial^{2}}{\partial x_{3}^{2}})$ & $u$   &   $(x_1, x_2, x_3) $       & $(\rho, c_p, k, P)$  & $t$      \\
 \hline
\end{tabular}%
}
\end{table}

The weighted-sum residual form of the PDE with TD interpolation can be written as:

\begin{equation}
    \int_\Omega\delta u^{TD}(\bm{x})\left[\mathcal{L}\left(u^{TD}(\bm{x})\right)-f(\bm{x})\right]d\Omega=0
\label{weighted_sum}
\end{equation}
where $u^{TD}$ is the approximation of the solution (i.e., trial function), $\delta u^{TD}$ is the test function, and $d\Omega =d\Omega_{\bm{x}_s}d\Omega_{\bm{x}_p}d\Omega_{{x}_t} $.

Depending on how $\delta u^{TD}$ is adopted, different mathematical formulations can be obtained. If the test function resides in the same function space as the trial function, it becomes the Galerkin formulation.  When the test function space differs from the trial function space, it becomes the Petrov-Galerkin formulation \cite{hughes2003finite}. If the Dirac delta function is used for the test function, then Eq. \ref{weighted_sum} corresponds to the collocation method \cite{reddy2005introduction}. In this paper, we employ the Galerkin formulation. However, the proposed framework is versatile and can be extended to accommodate other formulations as well.

\begin{figure}[!hbt]
\centering
\includegraphics[width=0.9\linewidth]{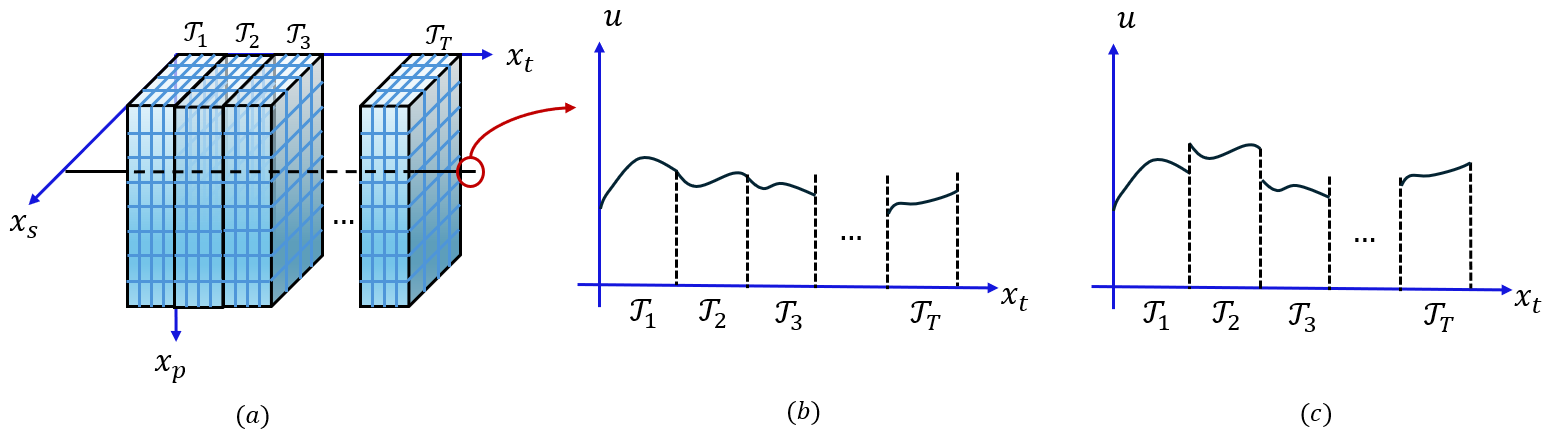}
\caption{(a) Multiple S-P-T slabs along the temporal dimension. (b) Continuous Galerkin: the solution is continuous across different S-P-T slabs. (c) Discontinuous Galerkin: jumps are allowed across the slab boundaries}
\label{spt}
\end{figure}

In Eq. \ref{td_eq_0}, the entire S-P-T domain is approximated using TD interpolation. However, this approach may result in a large system of equations due to the rapid increase in the number of TD modes for certain cases. For example, if the forcing function represents a moving source function in Eq. \ref{eq:pde}), this complexity may arise. To maintain computational efficiency, we can partition the temporal domain into a series of time slabs. As illustrated in Fig. \ref{spt}(a), the S-P-T continuum is divided into S-P-T slabs $\mathcal{T}_1, \mathcal{T}_2, \ldots, \mathcal{T}_T$. The solution within each time slab is then approximated individually using the TD interpolation.

Between consecutive S-P-T slabs, either a continuous or discontinuous formulation can be employed. As shown in Fig. \ref{spt}(b) for the continuous Galerkin scheme, the continuity of the solution in time is enforced by imposing the solution at the end of slab $\mathcal{T}_{i-1}$ as the initial condition of $\mathcal{T}_{i}$:

\begin{equation}
{}^{[\mathcal{T}+1]}{u}(\bm{x}_s,\bm{x}_p, 0) = {}^{[\mathcal{T}]}{u}(\bm{x}_s,\bm{x}_p, x_t^{max})
\label{slab}
\end{equation}
Discontinuous Galerkin method can be used when a discontinuity is allowed between S-P-T slabs, as illustrated in Fig. \ref{spt}(c). Discontinuity in time can be modeled using the jump operator $\llbracket...\rrbracket$ \cite{hughes1988space}.

\begin{equation}
    \llbracket {u}  (\bm{x}_s,\bm{x}_p, t)\rrbracket 
 =\lim_{\epsilon\to0^+}\left({}^{[\mathcal{T}+1]}{u}(\bm{x}_s,\bm{x}_p, \epsilon)-{}^{[\mathcal{T}]}{u}(\bm{x}_s,\bm{x}_p, x_t^{max} -\epsilon)\right)
\end{equation}

Keeping in mind that this approach can be applied generally to a range of engineering problems, we will demonstrate an example of the Galerkin formulation using a single space-parameter-time partition (S-P-T) slab in the remainder of this section. For illustrative purposes, the transient heat transfer equation will be utilized:
\begin{equation}
    \rho c_p \nabla _{x_t}u-\nabla_{\bm{x}_s}\cdot k \nabla_{\bm{x}_s}u=f(\bm{x}_s, \bm{x}_p,x_t)
\label{pde_eqn}
\end{equation}
as we focus on the example of modeling the laser powder bed fusion (LPBF) process in additive manufacturing (AM). In an LPBF simulation, we adopt the following time-dependent moving heat source function:
\begin{equation}
f(\bm{x}_s, \bm{x}_p,x_t) = \frac{2\eta P}{\pi r^2 d_v}\exp\left(-\frac{2\left(\left(x-x_0(t)\right)^2+(y-y_0(t))^2\right)}{r^2}\right)\cdot\mathbf{1}_{(x_3\geq d_v)}
\label{source}
\end{equation}
Summarizing the independent variables in Eq. \ref{pde_eqn}, there are spatial variables $\bm{x}_s = (x_1, x_2, x_3)$; parametric variables $\bm{x}_p = (k, \rho, c_p, \eta, P, r, d_v)$; and a temporal variable $x_t = t$. Among the parametric variables, $k$ is conductivity; $\rho$ is the material density; $c_p$ is heat capacity; $\eta$ is the material absorptivity; $P$ represents laser power; $r$ is the standard deviation that characterizes the width of the heat source; $d_v$ is the penetration depth of the heat source. In Eq. \ref{source}, $[x_0 (t),y_0 (t)]$ represents the center of the moving heat source; $\mathbf{1}_{(x_3\geq d_v)}$ is the indicator function where $\mathbf{1}_{(x_3\geq d_v)}=1$ if $x_3\geq d_v$ or $\mathbf{1}_{(x_3\geq d_v)}=0$ if $x_3<d_v$. Note that the discretization of the material parameters, in particular, in a random field setting, has been previously proposed by Liu et al. \cite{Liu1986Probabilistic,Liu1986Random}.

As shown in the schematic below, we classify the boundary surfaces into 2 categories: the Dirichlet boundary surface $\Gamma_D$ and the Neumann boundary surface $\Gamma_N$. 

\begin{figure}[!hbt]
\centering
\includegraphics[width=0.4\linewidth]{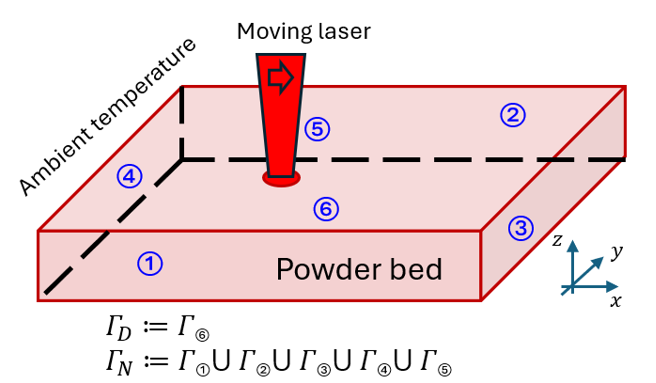}
\caption{Transient heat transfer with initial condition and boundary conditions.}
\label{am_boundary}
\end{figure}

A uniform ambient temperature is used as the initial condition. The bottom of the powder bed is subject to the Dirichlet boundary condition and the Neumann boundary conditions are prescribed on the other surfaces. The initial and boundary conditions are:
\begin{equation}
\begin{aligned}
    &u(\bm{x}_s, \bm{x}_p, 0)|_{\Omega} = u_0, \\
&u(\bm{x}_s, \bm{x}_p, x_t)|_{\Gamma_D} = u_0, \\
&\boldsymbol{n} \cdot \bm{q}|_{\Gamma_N} = q_{conv} + q_{rad} + q_{evap}
\end{aligned}
\end{equation}
where $u_0$ is the ambient temperature, $q_{conv}$ accounts for free convection, $q_{rad}$ accounts for radiation, and $q_{evap}$ imposes evaporative cooling when any material surface reaches the evaporation temperature \cite{liao2023efficient}. Each flux is defined as:

\begin{equation}
\begin{aligned}
&q_{conv} = h_{conv}[u(x,t) - u_0], \\
&q_{rad} = -\sigma_{SB} \epsilon (u^4(\bm{x}_s, x_t) - u_0^4), \\
&q_{evap} = -m_{evap} L_{evap}.
\end{aligned}
\end{equation}
where $\sigma_{SB}$ is the Stefan-Boltzmann constant; $\epsilon$ is the material emissivity; $u_0$ is the ambient temperature; $h_{conv}$ is the convection coefficient of the surrounding gas, $m_{evap}$ is the mass evaporation flux and $L_{evap}$ is the heat of evaporation. In the following numerical examples, we only consider the free convection term in the Neumann boundary condition.
The solution to Eq. \ref{pde_eqn} is approximated using TD interpolation function:
\begin{equation}
    u^{TD}(\bm{x}_s,\bm{x}_p,x_t)=\sum_{m=1}^M u_{\bm{x}_s}^{(m)}(\bm{x}_s)u_{\bm{x}_p}^{(m)}(\bm{x}_p)u_{x_t}^{(m)}(x_t)
\label{trial}
\end{equation}
Here a general notation is employed to represent different types of components in Eq. \ref{trial}. For example, the spatial component $u_{\bm{x}_s}^{(m)}(\bm{x}_s)$ is equivalent to $u_{x_1}^{(m)}(x_1) u_{x_2}^{(m)}(x_2) u_{x_3}^{(m)}(x_3)$. The corresponding test function can be obtained using the variational principle:

\begin{equation}
    \delta u^{TD}(\bm{x}_s,\bm{x}_p,x_t)=\sum_{m=1}^M\left[\delta u_{\bm{x}_s}^{(m)}(\bm{x}_s)u_{\bm{x}_p}^{(m)}(\bm{x}_p)u_{x_t}^{(m)}(x_t)+ u_{\bm{x}_s}^{(m)}(\bm{x}_s)\delta u_{\bm{x}_p}^{(m)}(\bm{x}_p)u_{x_t}^{(m)}(x_t)+u_{\bm{x}_s}^{(m)}(\bm{x}_s)u_{\bm{x}_p}^{(m)}(\bm{x}_p)\delta u_{x_t}^{(m)}(x_t)\right]
\label{test}
\end{equation}
Plugging the trial and test functions, the S-P-T Galerkin form of Eq. \ref{pde_eqn} can be obtained by following Eq. \ref{weighted_sum}:

\begin{equation}
    \int_\Omega\delta u^{TD}\left[\rho c_p \nabla _{x_t}u^{TD}-\nabla_{\bm{x}_s}\cdot k \nabla_{\bm{x}_s}u^{TD} - f\right]d\Omega=0
\label{weighted_sum1}
\end{equation}
Using integration by parts on the diffusion term, we get the corresponding general S-P-T Galerkin weak form in TAPS formulation:

\begin{equation}
    \int_{\Omega}\delta u^{TD} \rho c_p \nabla_{x_t} u^{TD} d\Omega +\int_{\Omega}\nabla_{\bm{x}_s}\delta u^{TD}\cdot k\nabla_{\bm{x}_s}u^{TD}d\Omega-\int_{\partial {\bm{x}_s},\bm{x}_p,t}\delta u^{TD}\bm{n}\cdot\bm{q}|_{\Gamma_N}d\bm{s}d\Omega_{\bm{x}_p} d\Omega_t-\int_{\Omega}\delta u^{TD} f(\bm{x}_s, \bm{x}_p,x_t) d\Omega=0
\label{galerkin}
\end{equation}
where $\bm{q}$ is the heat flux on the Neumann boundary.



\subsection{Discretized matrix form}

The S-P-T Galerkin weak form shown in Eq.  \ref{galerkin} is  nonlinear in nature due to the tensor product structure in TD interpolation, necessitating efficient solution schemes. To illustrate the detailed solution approach for the general S-P-T weak form, we simplify the governing equation Eq. \ref{pde_eqn} by considering a one-dimensional spatial problem where $x_s = x$. We assume that the product of density and specific heat capacity $\rho c_p$ is equal to 1. Additionally, the forcing term is solely dependent on $x$. Therefore, the simplified governing equation for this example is given by:

\begin{equation}
    \frac{\partial u}{\partial t}-\frac{\partial u}{\partial x}\cdot k \frac{\partial x}{\partial x}=f(x)
\label{pde_eqn1}
\end{equation}
subject to homogeneous boundary conditions and initial conditions. This equation has 3 independent variables ($D=3$), i.e., spatial variable $x_s = x_1=x$, parametric variable $x_p = x_2= k$ and temporal variable $x_t = x_3 =t$. The S-P-T Galerkin weak form of this problem can be written as follows according to Eq. \ref{galerkin} (the superscripts ``TD'' for both trial and test functions are omitted for brevity).

\begin{equation}
    \int_{\Omega}\delta u \nabla_{t} u d\Omega +\int_{\Omega}\nabla_{x}\delta u\cdot k\nabla_{x}ud\Omega - \int_{\Omega}\delta u f d\Omega = 0
\label{galerkin1}
\end{equation}
The corresponding trial and test functions can be obtained using Eqs. \ref{trial}-\ref{test}:
\begin{equation}
    u(x,k,t)=\sum_{m=1}^M u_{x}^{(m)}(x)u_{k}^{(m)}(k)u_{t}^{(m)}(t)
\label{td1}
\end{equation}

\begin{equation}
    \delta u(x,k,t)=\underbrace{\sum_{m=1}^M\delta u_{x}^{(m)}(x)u_{k}^{(m)}(k)u_{t}^{(m)}(t)}_\textrm{spatial variation}+\underbrace{\sum_{m=1}^M u_{x}^{(m)}(x)\delta u_{k}^{(m)}(k)u_{t}^{(m)}(t)}_\textrm{parametric variation}+\underbrace{\sum_{m=1}^Mu_{x}^{(m)}(x)u_{k}^{(m)}(k)\delta u_{t}^{(m)}(t)}_\textrm{temporal variation}
\label{test1}
\end{equation}
As shown in Eq. \ref{test1}, the test function is further split into $D$ variational terms for a general $D$ dimensional problem (in the current example, $D=3$). As an example, we first plug Eq. \ref{td1} and the spatial variation term of Eq. \ref{test1} into the Galerkin weak form in Eq. \ref{galerkin1} to obtain the S-P-T weak form terms corresponding to spatial variation:

\begin{equation}
    \begin{gathered}\underbrace{\int_{\Omega}\sum_{m=1}^M\sum_{n=1}^M \left[\nabla\delta u_{x}^{(m)}({x}) \nabla u_{x}^{(n)}({x})d{x}\right]\cdot \left[u_{k}^{(m)}(k) ku_{k}^{(n)}(k)d{k}\right]\cdot \left[u_t^{(m)}({t})u_{t}^{(n)}({t})d{t}\right]}_\textrm{diffusion term}+\\\underbrace{\int_{\Omega}\sum_{m=1}^{M}\sum_{n=1}^{M}\left[\delta u_{x}^{(m)}({x})u_{x}^{(n)}(x)d{x}\right]\cdot \left[u_{k}^{(m)}({k})u_{k}^{(n)}({k})d{k}\right]\cdot  \left[u_{t}^{(m)}({t}) 
 \nabla_{t}u_{t}^{(n)}({t})dt\right]}_\textrm{time derivative term} - \\\underbrace{\int_{\Omega}\sum_{m=1}^M\left[\delta u_{x}^{(m)}({x})f(x)d{x}\right]\cdot \left[u_{k}^{(m)}({k})d{k}\right]\cdot \left[u_{t}^{(m)}({t})d{t}\right]}_\textrm{forcing term}\end{gathered}
\label{subspace_x_expand_simple}
\end{equation}
We use 1D C-HiDeNN shape functions to approximate each univariate function: 

\begin{eqnarray}
 u_d^{(n)}(x_d) = \widetilde{N}^{[d]}_{n_d'}(x_d)u_{n_d'n}^{[d]} \quad  (\text{no sum on $d$}) \nonumber\\ 
 \delta u_d^{(m)}(x_d) = \widetilde{N}^{[d]}_{n_d}(x_d)\delta u_{n_dm}^{[d]} \quad   (\text{no sum on $d$}) 
\label{shape}
\end{eqnarray}
where Einstein summation is used. The free index $d$ refers to dimension and $d=x,k$ or $t$. The gradient of the interpolated variable can be computed using the shape function derivative $\widetilde{B}^{[d]}_{n_d}(x_d) = \frac{d \widetilde{N}^{[d]}_{n_d}(x_d)}{d x_d}$.

\begin{eqnarray}
 \nabla_{x_d}u_d^{(n)}(x_d) =   \widetilde{B}^{[d]}_{n_d'}(x_d)  u_{n_d'n}^{[d]} \quad   (\text{no sum on $d$}) \nonumber\\ 
 \nabla_{x_d}\delta u_d^{(m)}(x_d) =  \widetilde{B}^{[d]}_{n_d}(x_d)  \delta u_{n_dm}^{[d]}   \quad   (\text{no sum on $d$})
\label{shape_deri}
\end{eqnarray}
Plugging Eq. \ref{shape} - \ref{shape_deri} into Eq. \ref{subspace_x_expand_simple}, the diffusion term can be rewritten as: 

\begin{equation}
\sum_{m=1}^M\sum_{n=1}^M \underbrace{\int_{\Omega_x} \widetilde{B}_{n_x}(x)  \delta u_{n_xm}^{[x]} \widetilde{B}_{n_x'}(x)  u_{n_x'n}^{[x]} dx}_\textrm{spatial term}\cdot \underbrace{\int_{\Omega_k}\widetilde{N}_{n_k}(k)u_{n_km}^{[k]} k\widetilde{N}_{n_k'}(k)u_{n_k'n}^{[k]}d{k}}_\textrm{parametric term}\cdot \underbrace{\int_{\Omega_t}\widetilde{N}_{n_t}(t)u_{n_tm}^{[t]}\widetilde{N}_{n_t'}(t)u_{n_t'n}^{[t]}d{t}}_\textrm{temporal term}
\label{subspace_x_expand_diffusion}
\end{equation}
As can be readily seen from Eq. \ref{subspace_x_expand_diffusion}, after doing 1D integration of each term, the parametric and temporal terms can be treated as coefficient matrices:

\begin{eqnarray}
 C^{[k]}_{mn} = \underbrace{\int_{\Omega_k}\widetilde{N}_{n_k}(k)u_{n_km}^{[k]} k\widetilde{N}_{n_k'}(k)u_{n_k'n}^{[k]}d{k}}_\textrm{parametric term}  \nonumber\\ 
 C^{[t]}_{mn} =  \underbrace{\int_{\Omega_t}\widetilde{N}_{n_t}(t)u_{n_tm}^{[t]}\widetilde{N}_{n_t'}(t)u_{n_t'n}^{[t]}d{t}}_\textrm{temporal term}
\label{eq:coef}
\end{eqnarray}
as the only free indices are $m$ and $n$. Substituting the coefficient matrices and rearranging different terms in Eq. \ref{subspace_x_expand_diffusion}, we have:

\begin{equation}
\sum_{m=1}^M \delta u_{n_xm}^{[x]} \sum_{n=1}^M  \left[\int_{\Omega_x}\widetilde{B}_{n_x}(x) \widetilde{B}_{n_x'}(x) dx\right]  \cdot C^{[k]}_{mn} C^{[t]}_{mn} \cdot u_{n_x'n}^{[x]}
\label{subspace_x_expand_diffusion1}
\end{equation}
Like standard FEM, we can define $\int_{x}\widetilde{B}_{n_x}(x) \widetilde{B}_{n_x'}(x) dx$ as the 1D stiffness matrix $K^{[x]}_{n_xn_x'}$ of $x$ dimension in Eq. \ref{subspace_x_expand_diffusion1}. We let $C^{[x]}_{mn} = C^{[k]}_{mn} C^{[t]}_{mn}$ with no summation on $(m,n)$. Furthermore, let us define the following 4-th order tensor:

\begin{equation}
A_{n_xn_x'mn}^{[x]} =  K^{[x]}_{n_xn_x'} C^{[x]}_{mn}
\label{a_tensor}
\end{equation}
where $A_{n_xn_x'mn}^{[x]}$ is a function of solution vectors $u_{n_km}^{[k]}$ and $u_{n_tm}^{[t]}$ since the coefficient matrix $C^{[x]}_{mn}$ depends on these solution vectors as shown in Eq. \ref{eq:coef}. This dependency reflects the interconnected nature of the variables across different dimensions in the S-P-T framework, highlighting how the spatial, parameter, and temporal components influence each other through the coefficients. As a result, Eq. \ref{subspace_x_expand_diffusion} can be further simplified as follows:
\begin{equation}
 \delta u_{n_xm}^{[x]}  A_{n_xn_x'mn}^{[x]} u_{n_x'n}^{[x]}
\label{subspace_x_expand_diffusion2}
\end{equation}
where the summation signs are neglected since $m$ and $n$ become dummy variables. The 4-th order tensor $A_{n_xn_x'mn}^{[x]}$ can be reshaped as a 2nd order tensor $\mathbb{A}_{IJ}^{[x]}$: the indices $ n_x $ and $ m $ are combined into a single composite index $ I $, and the indices $ n_x' $ and $ n $ are combined into a single composite index $ J $. 

\begin{equation}
    A_{n_xn_x'mn}^{[x]} =  \mathbb{A}_{IJ}^{[x]}
\end{equation}
Define the following vectorization:

\begin{eqnarray}
 \delta \mathbb{U}^{[x]}_I = \left[\text{vec}\left(\delta u_{n_xm}^{[x]}\right)\right]_I \nonumber\\ 
  \mathbb{U}^{[x]}_J = \left[\text{vec}\left(u_{n_x'n}^{[x]}\right)\right]_J
\end{eqnarray}
As a result, Eq. \ref{subspace_x_expand_diffusion2} is equivalent to:

\begin{equation}
\delta \mathbb{U}^{[x]^T}  \mathbb{A}^{[x]} \mathbb{U}^{[x]}
\label{subspace_x_expand_diffusion3}
\end{equation}
Following the same procedure, we can obtain matrix forms corresponding to the time derivative term $\delta \mathbb{U}^{[x]^T} \mathbb{B}^{[x]}\mathbb{U}^{[x]}$, and the forcing term $\delta \mathbb{U}^{[x]^T}\mathbb{Q}^{[x]}$ for the spatial variational part of Eq. \ref{subspace_x_expand_simple}. Similar structures can also be obtained for the parametric and temporal variational parts of the test function in Eq. \ref{test1}. Denoting $\mathbb{K}^{[d]} = \mathbb{A}^{[d]} + \mathbb{B}^{[d]}$, the matrix form of the generalized S-P-T Galerkin form in Eq. \ref{galerkin1} can be written as:
\begin{equation}
\underbrace{\delta\mathbb{U}^{[x]^T}\mathbb{K}^{[x]}\mathbb{U}^{[x]}- \delta \mathbb{U}^{[x]^T}\mathbb{Q}^{[x]}}_\textrm{spatial variational part} + \underbrace{\delta\mathbb{U}^{[k]^T}\mathbb{K}^{[k]}\mathbb{U}^{[k]} - \delta \mathbb{U}^{[k]^T}\mathbb{Q}^{[k]}}_\textrm{parametric variational part} +  \underbrace{\delta\mathbb{U}^{[t]^T}\mathbb{K}^{[t]}\mathbb{U}^{[t]} -
\delta \mathbb{U}^{[t]^T}\mathbb{Q}^{[t]}}_\textrm{temporal variational part}   = 0
\label{part1}
\end{equation}
Eq. \ref{part1} is equivalent to the following nonlinear system of equations. Note that the nonlinearity comes from the fact that $\mathbb{K}^{[d]}$ is solution dependent:

\begin{equation}
\left[\delta \mathbb{U}^{[x]^T}, \delta \mathbb{U}^{[k]^T}, \delta \mathbb{U}^{[t]^T} \right] \left\{\begin{bmatrix}
\mathbb{K}^{[x]}(\mathbb{U}^{[k]},\mathbb{U}^{[t]}) & 0 & 0        \\
0      &\mathbb{K}^{[k]}(\mathbb{U}^{[x]},\mathbb{U}^{[t]})  & 0  \\
0      &0       & \mathbb{K}^{[t]}(\mathbb{U}^{[x]},\mathbb{U}^{[k]})
\end{bmatrix}
\begin{bmatrix}
\mathbb{U}^{[x]} \\
\mathbb{U}^{[k]} \\
\mathbb{U}^{[t]} 
\end{bmatrix}
 -\begin{bmatrix}
\mathbb{Q}^{[x]}(\mathbb{U}^{[k]},\mathbb{U}^{[t]}) \\
\mathbb{Q}^{[k]}(\mathbb{U}^{[x]},\mathbb{U}^{[t]}) \\
\mathbb{Q}^{[t]}(\mathbb{U}^{[x]},\mathbb{U}^{[k]}) 
\end{bmatrix}\right\}=0
\label{part3}
\end{equation}
where we can treat the solution vector $\left[ \mathbb{U}^{[x]^T},  \mathbb{U}^{[k]^T},  \mathbb{U}^{[t]^T} \right]$ as generalized DoFs like standard FEM. There are many ways to solve Eq. \ref{part3}. For example, standard linearization schemes such as Newton's method have been used \cite{ammar2006new}. However, this method may suffer from ill-conditioning since the mismatch of scales for different dimensions can be significant. In this paper, we use the concept of subspace iteration to efficiently approximate the solution by iterating in the subspace of the test function space until a convergence criteria is met \cite{nouy2010priori}. A similar counterpart has been widely adopted as the gold standard in discrete tensor decomposition \cite{kolda2009tensor}.



\subsection{Solution scheme of TAPS: subspace iteration}
For subspace iteration in $d$-th dimension, only the solution matrix $\mathbb{U}^{[d]}$ is treated as unknown while all other functions are considered as known constants. Consequently, the variations of the univariate functions other than $d$-th dimension will vanish. From Eq. \ref{part3}, it can be seen that this will lead to a linear system of equations for the unknowns in the $d$-th dimension. The updated solution matrix $\mathbb{U}^{[d]}$ from this process is then used in the next subspace iteration for dimension $d+1$ (when $d=D$, we come back to the first dimension $d=1$). The complete solution scheme for subspace iteration is shown in Algorithm 1.

\begin{algorithm}
\caption{TAPS solution scheme (subspace iteration)}
\label{alg:linear_td}
\begin{algorithmic}[1]
\State Initialize solution vector $\mathbb{U}^{[x_1][0]},...,\mathbb{U}^{[x_D][0]}$ with random values and compute $\mathbb{K}^{[x_1][{0}]}$,  and $\mathbb{Q}^{[x_1][{0}]}$
\For{$iter = 0$ to $iter_{max}$}
    \For{$d = 1$ to D}
        \State Update iteration number $\mathcal{K} = iter \times D + d$,
        \State Solve TD linear system $\mathbb{K}^{[x_d][\mathcal{K}-1]}   \mathbb{U}^{[x_d][\mathcal{K}]} = \mathbb{Q}^{[x_d] [\mathcal{K}-1]} $  
        \State Update matrices $\mathbb{K}^{[x_{d+1}][\mathcal{K}]}$ and force vector $\mathbb{Q}^{[x_{d+1}][\mathcal{K}]}$

    \EndFor
    \State Check convergence
\EndFor
\end{algorithmic}
\end{algorithm}

To illustrate the details of the subspace iteration algorithm, we consider the $\mathcal{K}$-th subspace iteration (which is on spatial variable $x$). Here, we assume that the parametric and temporal solutions have been updated from the previous $(\mathcal{K} - 1)$-th iteration, leaving the spatial solution as unknown to be solved in $\mathcal{K}$-th iteration. Moreover, instead of the full variation form of the test function as in Eq. \ref{part3}, we only consider the subspace $x$ of the test function by setting the parametric and temporal variational parts as 0. As a result, we have:


\begin{equation}
\mathbb{K}^{[x][\mathcal{K}-1]} \left(\mathbb{U}^{[k][\mathcal{K}-1]},\mathbb{U}^{[t][\mathcal{K}-1]}\right) 
\mathbb{U}^{[x][\mathcal{K}]}  = \mathbb{Q}^{[x][\mathcal{K}-1]}\left(\mathbb{U}^{[k][\mathcal{K}-1]},\mathbb{U}^{[t][\mathcal{K}-1]}\right)
\label{part5}
\end{equation}
which is a linear system of equations with unknown $\mathbb{U}^{[x][\mathcal{K}]}$. This is a general Sylvester equation which can be solved using many efficient solution schemes \cite{bouhamidi2008note, xie2015scaling}. In this paper, sparse direct solvers based on fast diagonalization/complex Schur decomposition methods are adopted  \cite{langer2021efficient}. The computational complexity of the sparse direct solver is  $\mathcal{O}(M^3+M^2 n_d+C_c (n_d))$ for the $d$-th dimension subspace iteration, where $M$ is the total number of modes; $n_d$ is the number of grid points for $d$-th dimension; $C_c (n_d)$ refers to the computational cost of the banded sparse mass/stiffness matrix for $d$-th dimension with a shape of $(n_d\times n_d)$. 

Once $\mathbb{U}^{[x][\mathcal{K}]}$ is obtained, we then update matrix $\mathbb{K}^{[k][\mathcal{K}]}(\mathbb{U}^{[x][\mathcal{K}]},\mathbb{U}^{[t][\mathcal{K}]}) $ and forcing vector $\mathbb{Q}^{[k][\mathcal{K}]}(\mathbb{U}^{[x][\mathcal{K}]},\mathbb{U}^{[t][\mathcal{K}]}) $. In the next iteration (for dimension $k$), we treat $\mathbb{U}^{[k][\mathcal{K} + 1]}$ as the only unknown. Subspace iteration will continue unless the relative change of all solution matrices (for example, $L_2$ norm) is within the tolerance. 

\subsection{Error estimates of TAPS}
Since the TAPS solution is based on the C-HiDeNN-TD approximation and the generalized Galerkin formulation, we can have the following theoretical results on the error bounds, as demonstrated in our previous work on C-HiDeNN \cite{lu2023convolution}:
\begin{equation}
\label{eq:accuracy}
 \begin{aligned}     
     \|u^{\text{C-HiDeNN}}-u^{\text{ex}}\|_E 
    \leq \|u^{\text{TAPS}}-u^{\text{ex}}\|_E 
    \leq \|u^{\text{FEM}}-u^{\text{ex}}\|_E 
     \end{aligned}
\end{equation}
where  $\|\cdot\|_E $ denotes the energy norm, $u^{\text{ex}}$ denotes the exact solution, $u^{\text{C-HiDeNN}}$ denotes the solution obtained by  the full C-HiDeNN method without tensor decomposition, $u^{\text{TAPS}}$ denotes the TAPS solution with a sufficient number of modes, $u^{\text{FEM}}$ denotes the FEM solution. The proof of the above results is based on the fact that the full C-HiDeNN approximation can provide a larger function space and therefore more accurate solutions than conventional FEM \cite{lu2023convolution}. The subspace iteration can be considered as a local (directional) version of the Galerkin formulation and is expected to enable an optimized solution for the tensor decomposition that  will converge to the Galerkin-based full C-HiDeNN method.

\section{Results}
\subsection{Convergence study for moving heat source}
In this section, we first analyze the convergence of the TAPS solver for a space-time (S-T) transient heat transfer problem. A single NVIDIA RTX A6000 GPU is used for all the following analyses. In Eq. \ref{pde_eqn}, we let $\rho c_p=1$, $k = 1$, and replace the heat source as shown in Eq. \ref{source_eq1}. In this example, we have the spatial variable $\bm{x}_s = (x,y,z)$ and the temporal variable $x_t = t$. 

\begin{equation}
    \begin{aligned}f(\bm{x}_s,x_t)=2(1&-2y^2)(1-e^{-15t})e^{-y^2-(100t-x-5)^2}\\&+2(1-2(100t-x-5)^2)(1-e^{-15t})e^{-y^2-(100t-x-5)^2}+(1\\&-e^{-15t})(200x+1000-20000t)e^{-y^2-(100t-x-5)^2}\\&-15e^{-15t}e^{-y^2-(100t-x-5)^2}\end{aligned}
\label{source_eq1}
\end{equation}
The analytical solution to the PDE is inherently non-separable. 
\begin{equation}
    u^{\text{ex}}(\bm{x}_s,x_t)=(1-e^{-15t})e^{-y^2-(x-100t-5)^2}
\end{equation}
The initial and boundary conditions are:

\begin{equation}
\begin{aligned}
& u(\bm{x}_s, 0) = 0, \\
& u(\bm{x}_s, x_t)\big|_{\partial\Omega} = u^{\text{ex}}(\bm{x}_s, x_t)\big|_{\partial\Omega}.
\end{aligned}
\end{equation}
The relative ${L_2}$ norm error is defined as:
\begin{equation}
    \epsilon_{L_2}=\frac{\|u^{TD}(\bm{x}_s,x_t)-u^{\text{ex}}(\bm{x}_s, x_t)\|_{L_2 (\Omega_{\bm{x}_s} \otimes\Omega_{x_t})}}{\|u^{\text{ex}}(\bm{x}_s, x_t)\|_{L_2 (\Omega_{\bm{x}_s} \otimes\Omega_{x_t})}}
\label{eq:l2norm}
\end{equation}

\begin{figure}[!hbt]
\centering
\includegraphics[width=0.8\linewidth]{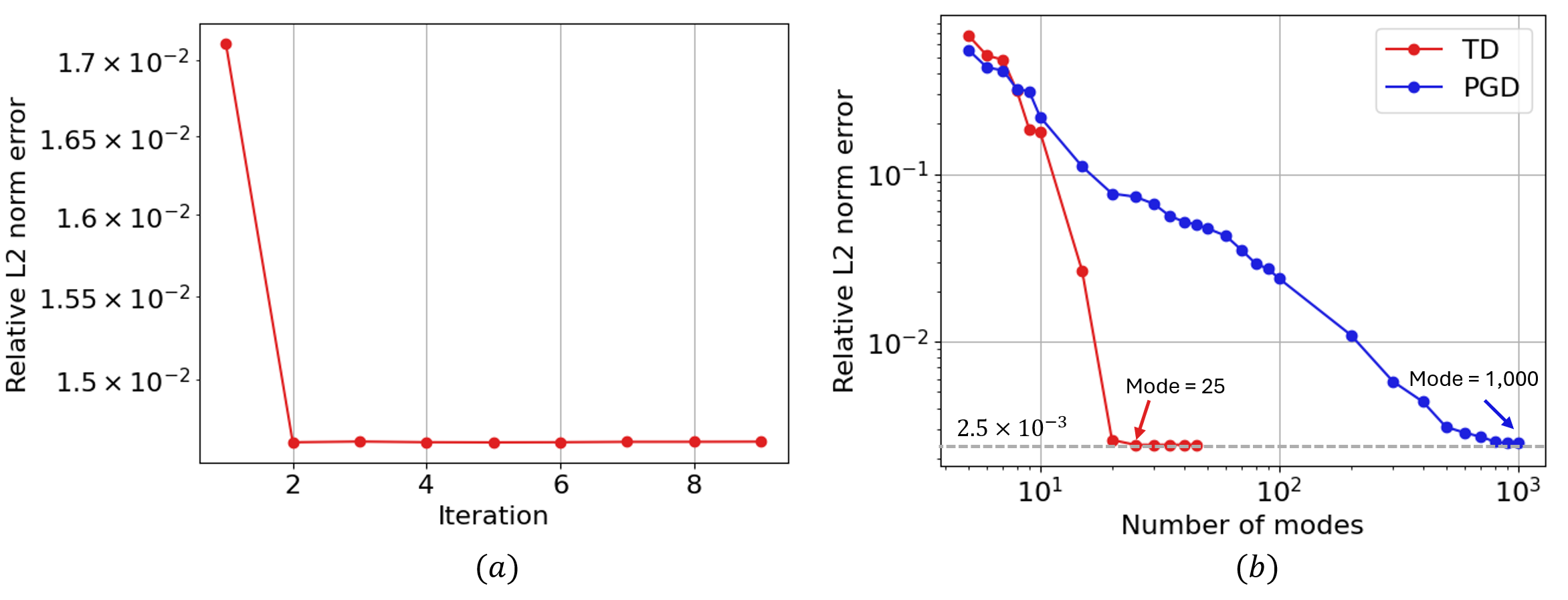}
\caption{Relative L2 norm error with respect to (a) the number of subspace iterations  (b) the number of modes}
\label{moving1}
\end{figure}

First, we investigate the influence of the number of subspace iterations. As shown in Fig. \ref{moving1}(a), 3 iterations are enough to obtain an accurate result. Next, we investigate the convergence in terms of the number of modes. Here we compare the relative ${L_2}$ norm error for both TAPS and proper generalized decomposition (PGD) methods \cite{chinesta2011overview, chinesta2013proper}. To this aim, we use the same discretization for the space-time domain with each dimension discretized by 100 grid points, the same reproducing polynomial order $p=1$ and convolution patch size $s=1$. As can be seen from Fig. \ref{moving1}(b), TAPS requires a much smaller number of modes than PGD. For TAPS, when the number of modes equals 25, the relative $L_2$ norm error decreases to $2.5\times10^{-3}$. The total solution time is 15.2 s. However, PGD requires 1,000 modes which takes 60.6 s solution time to reach the same level of accuracy. This is because the test function space in PGD is a subspace of TAPS \cite{ammar2006new}. Furthermore, the modal decomposition obtained from PGD is not optimal and thus requires a larger storage requirement due to the increased number of modes.

\begin{figure}[!hbt]
\centering
\includegraphics[width=1\linewidth]{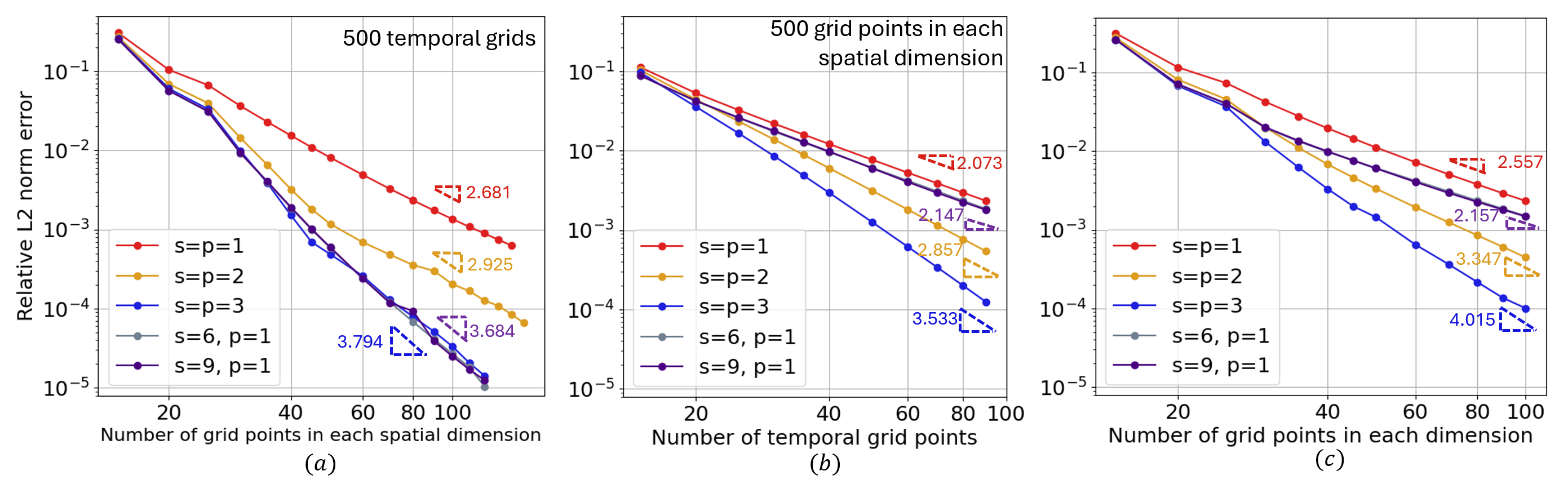}
\caption{Relative $L_2$ norm error with respect to the number of grid points (a) spatial convergence (b) temporal convergence (c) spatio-temporal convergence}
\label{moving2}
\end{figure}

The spatial and temporal convergence are also studied. In Fig. \ref{moving2}(a), the number of temporal nodes is fixed as 500, and the spatial mesh is refined. It shows the relative $L_2$ norm error with respect to the number of nodes along each spatial dimension. As can be readily seen from the figure, larger patch size $s$ leads to smaller error given the same reproducing polynomial orders $p$ . Moreover, we can adjust $p$ to control the spatial convergence rate. Similarly, Fig. \ref{moving2}(b) demonstrates the convergence rate in the temporal domain where we fix the spatial discretization as 500 along each spatial dimension. By adjusting $s$ and $p$, we can obtain different temporal convergence rates.

Finally, we refine the spatial and temporal mesh simultaneously and study the spatio-temporal convergence rate in Fig. \ref{moving2}(c). As can be observed from the figure, higher reproducing polynomial order $p$ will lead to a higher-order convergence rate. 

\subsection{Convergence study of S-P-T problems up to equivalent zetta-scale ($10^{21}$) full models}
\begin{figure}[!hbt]
\centering
\includegraphics[width=0.4\linewidth]{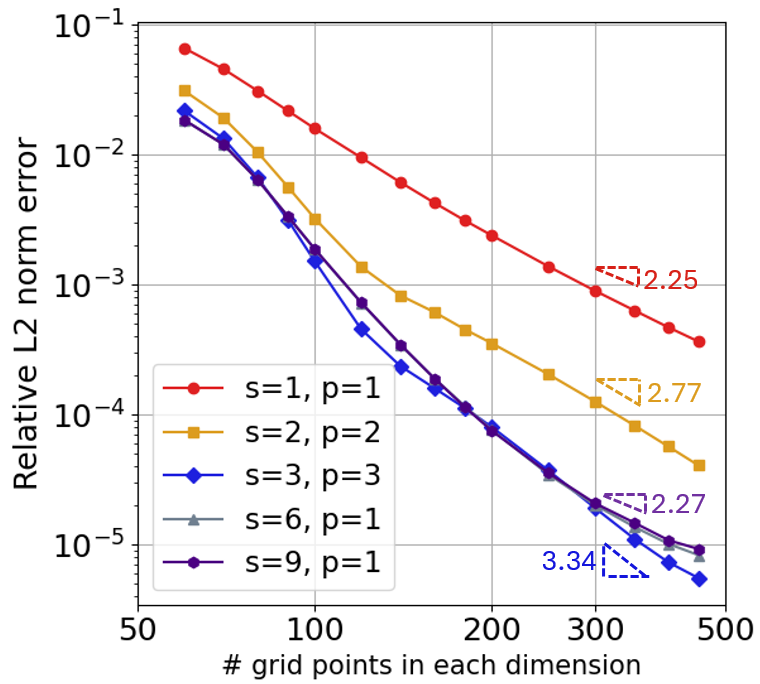}
\caption{Relative $L_2$ norm error with respect to the number of grid points in each dimension}
\label{SPT_convergence}
\end{figure}

In this example, we study the convergence of the TAPS solver for the time-dependent parametric heat transfer problem in a S-P-T setting. In Eq. \ref{pde_eqn}, we adopt the heat source as shown in Eq. \ref{SPT_source}. In this example, we have spatial variable $\bm{x}_s = (x,y,z)$, parametric variable $\bm{x}_p = (k, P, \rho, c_p)$ and temporal variable $x_t = t$. 

\begin{equation}
 \begin{aligned}
        f(\bm{x}_s,\bm{x}_p, x_t) = &15\rho^2 c_p^2kp e^{-15kt} e^{-25.0x^2 - 25.0y^2} \\
        & + 50\rho c_pkp\left(1 - e^{-15kt}\right) e^{-25.0x^2 - 25.0y^2} \left[ \left(1 - 50x^2\right) +\left(1 - 50y^2\right)  \right]
    \end{aligned}
\label{SPT_source}
\end{equation}
The analytical solution to the PDE is inherently non-separable. 
\begin{equation}
    u^{\text{ex}}(\bm{x}_s,\bm{x}_p, x_t)=\rho c_pP(1-e^{-15kt})e^{-25.0x^2 - 25.0y^2}
\end{equation}
The initial and boundary conditions are:

\begin{equation}
\begin{aligned}
& u(\bm{x}_s,\bm{x}_p, 0) = 0  \nonumber\\ 
& u(\bm{x}_s,\bm{x}_p, x_t)|_{\partial\Omega}=u^{\text{ex}}\left(\bm{x}_s, \bm{x}_p, x_t\right)|_{\partial\Omega} 
\end{aligned}
\end{equation}
The relative ${L_2}$ norm error is defined as:
\begin{equation}
    \epsilon_{L_2}=\frac{\|u^{TD}(\bm{x}_s,\bm{x}_p, x_t)-u^{\text{ex}}(\bm{x}_s,\bm{x}_p, x_t)\|_{L_2 (\Omega_{\bm{x}_s}\otimes \Omega_{\bm{x}_p}\otimes\Omega_{x_t})}}{\|u^{\text{ex}}(\bm{x}_s,\bm{x}_p, x_t)\|_{L_2 (\Omega_{\bm{x}_s}\otimes \Omega_{\bm{x}_p}\otimes\Omega_{x_t})}}
\label{eq:l2norm_spt}
\end{equation}
To study the convergence of TAPS for S-P-T problems, the number of grid points is refined simultaneously in each dimension and corresponding relative $L_2$ norm errors are computed as shown in Fig. \ref{SPT_convergence}. When the number of grid points in each dimension is 450, the equivalent DoFs of a full model achieves $450^8 = 1.68\times 10^{21}$. Consequently, it is equivalent to a zetta-scale ($10^{21}$) full problem. As can be seen from the figure, a larger patch size $s$ leads to a smaller error and faster convergence. A higher reproducing polynomial order $p$ also leads to a higher convergence rate. It can be noticed that the convergence rate for $p=3$ case is smaller than expected $p+1=4$. This is attributed to the fact that the S-P-T mesh is not fine enough. However, due to the rounding error in computing the relative $L_2$ norm error, we can only accurately compute the error up to 450 grid points per dimension.

In summary, we have the flexibility to choose different $s$ and $p$ to control the accuracy of TAPS by directly solving the S-P-T PDE. This is different from other data-driven modeling approaches (for instance, neural networks-based data-driven methods) in two notable ways. First, unlike a black-box neural network interpolator where the accuracy of the model is not guaranteed, our method is built upon the AI-enhanced finite element method, and we can control the convergence rate by choosing suitable hyperparameters $s$ and $p$. Second, unlike most data-driven reduced-order models for physical problems, our method directly solves the governing PDE by plugging in the TD interpolation without seeing any training data. As a result, we can avoid the most expensive offline data generation stage as opposed to data-driven methods.

\subsection{Moving source with solution dependent material parameters}
\begin{figure}[!hbt]
\centering
\includegraphics[width=0.8\linewidth]{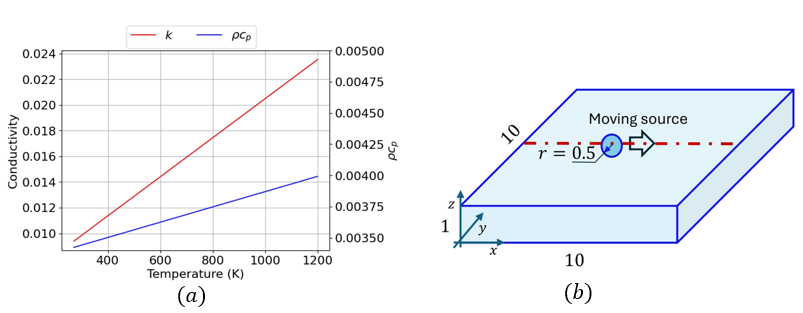}
\caption{(a) Temperature dependent material properties for Inconel 718 \cite{agazhanov2019thermophysical} (b) Schematic of numerical simulation, where the solution along the center line is compared for FEM and TAPS.}
\label{material}
\end{figure}

In this section, we model moving heat sources using temperature-dependent material parameters. The solution scheme of this problem is provided in detail in \ref{app:_nonlinear_PDE}. Figure \ref{material}(a) illustrates a typical representation of temperature-dependent heat conductivity and capacity for Inconel 718 \cite{agazhanov2019thermophysical}.

Since the temperature dependency of $k(u)$ and $\rho c_p (u)$ can be approximated using a linear relationship. As a result, we can directly rewrite $k(u(\bm{x}_s,x_t))$ and $\rho c_p (u(\bm{x}_s,x_t))$ in the TD format.

\begin{eqnarray}
k(\bm{x}_s,x_t) \approx \sum_{m=1}^M m_k u_{x_1}^{(m)}(x_1)u_{x_2}^{(m)}(x_2)u_{x_3}^{(m)}(x_3)u_{x_t}^{(m)}(x_t)+n_k  \nonumber\\ 
\rho c_p(\bm{x}_s,x_t) \approx \sum_{m=1}^M m_{c_p} u_{x_1}^{(m)}(x_1)u_{x_2}^{(m)}(x_2)u_{x_3}^{(m)}(x_3)u_{x_t}^{(m)}(x_t)+n_{c_p} 
\end{eqnarray}
where $M$ is the decomposition modes of the TAPS solution;  $m_k=1.52\times10^{-5}$  W/(mmK$^2$); $n_k=5.29\times10^{-3}$  W/(mmK); $m_{c_p }=6.11\times10^{-7}$ mm$^{-3}$K$^{-2}$; $n_{cp}=3.25\times10^{-3}$ mm$^{-3}$K$^{-1}$;

The problem setup is shown in Fig. \ref{material} (b). The spatial domain size is 10mm$\times$10mm$\times$1mm where homogeneous Dirichlet boundary conditions are assumed for the left and right surfaces; homogeneous Neumann boundary conditions are applied to all other surfaces. As shown in Eq. \ref{source}, a moving Gaussian source term $f(\bm{x}_s,t)$  is applied as a volumetric source term with a radius $r=0.5$ mm and moving velocity $500$ mm/s. The diameter is discretized using 10 spatial elements.

\begin{figure}[!hbt]
\centering
\includegraphics[width=0.4\linewidth]{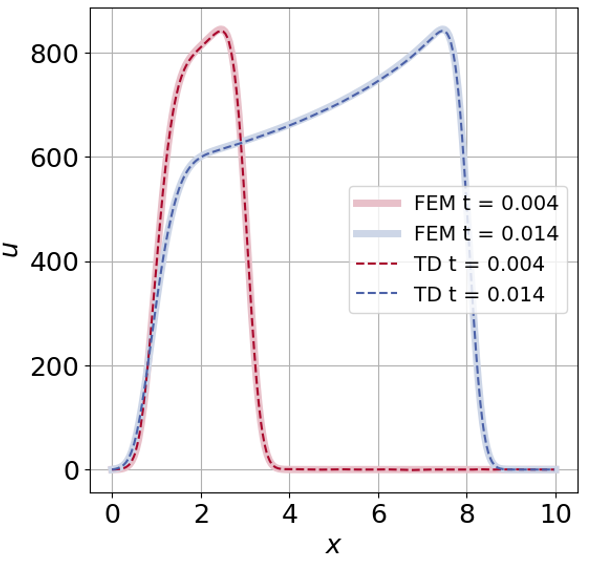}
\caption{Comparison of Nonlinear TAPS solution versus finite element solution at different times.}
\label{material_compare}
\end{figure}

Since there is no analytical solution available to this problem, we use implicit finite element analysis as the baseline for validation. JAX-FEM \cite{xue2023jax} is used to generate the nonlinear FEM solution. For ease of comparison, we use the same time increment as $1.60\times 10^{-4}$ sec for both TAPS and FEM. The solution along the center line, as shown in Fig. \ref{material} (b) is compared. As can be seen from Fig. \ref{material_compare}, the result of the nonlinear TAPS solver agrees well with FEM.

\subsection{Simulation of LPBF process}
In this section, we use TAPS to efficiently model the laser powder bed fusion process (LPBF) in additive manufacturing. Here we only consider the free convection term in the Neumann boundary condition.
The initial condition can be considered by splitting the total solution as a summation of the homogeneous part and the inhomogeneous part.

\begin{equation}
    u(\bm{x}_s, x_t)=u_0(\bm{x}_s, x_t)+u_{init}(\bm{x}_s)
\end{equation}
As a result, $u_0 (\bm{x}_s, x_t) $ is subject to homogeneous initial conditions.
In this section, we assume Ti-6Al-4V is used as the powder bed materials. The detailed material parameters can be found in Table \ref{tab:ti64}.

\begin{table}[!hbt]
\centering
\caption{Parameters used in the simulation}
\label{tab:ti64}
\begin{tabularx}{\textwidth}{l|X|X|X}
\hline
\textbf{Parameter} & \textbf{Variable} & \textbf{Value} & \textbf{Units} \\
\hline
Thermal conductivity & \( k \) & 22.0 & W m\(^{-1}\) K\(^{-1}\) \\
Density & \( \rho \) & 4.27 & g cm\(^{-3}\) \\
Specific heat capacity & \( c_p \) & 745 & J kg\(^{-1}\) K\(^{-1}\) \\
Ambient temperature & \( T_0 \) & 298.15 & K \\
Heat convection coefficient & \( h_{conv} \) & 14.73 & W m\(^{-2}\) K\(^{-1}\) \\
\hline
\end{tabularx}
\end{table}

\subsubsection{Single-track simulation}
\begin{figure}[!hbt]
\centering
\includegraphics[width=0.9\linewidth]{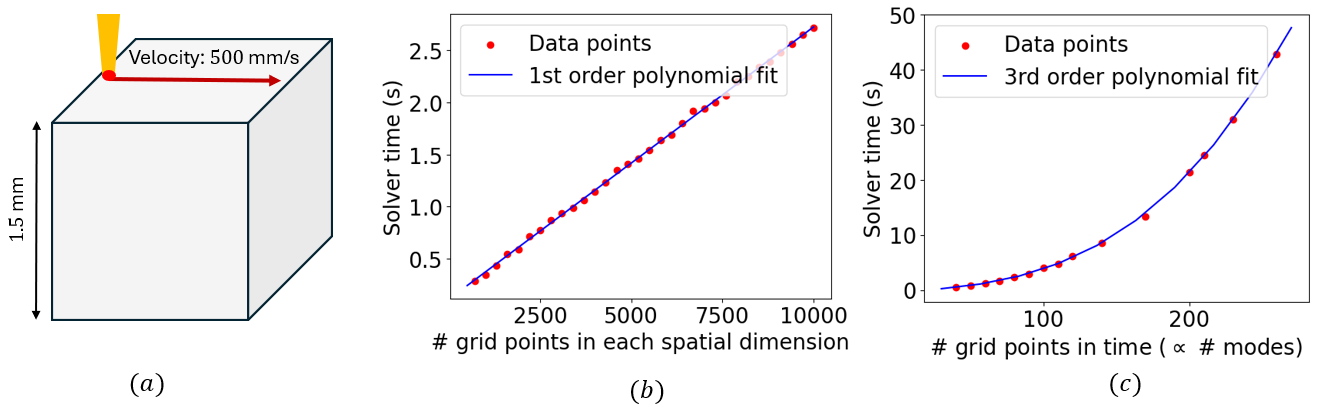}
\caption{(a) Single-track simulation. (b) Computational time of subspace iteration in refining the spatial mesh: linear growth. (c) Computational time of subspace iteration in refining the temporal mesh in a single space-time slab: cubic growth due to the increased number of modes}
\label{single_track}
\end{figure}

In this example, we investigate the computational complexity numerically for single-track LPBF simulation with a single S-T slab, as shown in Fig. \ref{single_track} (a). A single NVIDIA RTX A6000 GPU is used for all the following analyses. To ensure accuracy, the number of modes is adopted as 5 times larger than the number of time steps in the following examples. In the first case, within the S-T slab, the spatial mesh is refined uniformly along each spatial dimension while fixing the number of temporal grid points. The computational time for each subspace iteration is plotted in Fig. \ref{single_track} (b). It can be seen that TAPS has a linear growth of computational complexity when refining the spatial mesh.  

Similarly, we only refine the temporal mesh while fixing the spatial mesh in the second case and plot the computational time for each subspace iteration as in Fig. \ref{single_track} (c). It can be readily observed that refining the temporal mesh has a much higher computational complexity than refining the spatial mesh. This is because increasing temporal elements will also lead to an increased number of modes $M$. As mentioned before, the computational cost for the sparse direct solver employed is $\mathcal{O}(M^3+M^2 n_d+C_c (n_d))$ for the $d$-th dimension subproblem, where $M$ represents total number of modes; $n_d$ refers to the total number of grid points in $d$-th dimension; $C_c (n_d)$ refers to the computational cost of a banded sparse matrix with a shape of $(n_d\times n_d)$. Therefore, the increased number of modes leads to a cubic growth in computational time.

\begin{table}[!hbt]
\centering
\caption{Parameters used in the single-track simulation}
\label{tab:complexity}
\begin{tabularx}{\textwidth}{l|X|X|X}
\hline
\textbf{Parameter} & \textbf{Variable} & \textbf{Value} & \textbf{Units} \\
\hline
Laser power & \( P \) & 200 & W \\
Laser spot size radius & \( r \) & 50 & $\mu$m \\
Laser scan speed & \( V \) & 500 & mm s\(^{-1}\) \\
Absorptivity & \( \eta \) & 0.25 & 1 \\
Length & \( L \) & 1.5 & mm \\
Width & \( W \) & 1.5 & mm \\
Height & \( H \) & 1.5 & mm \\
Laser penetration depth & \( d \) & 50 & $\mu$m \\
Mesh size & \( h \) & 5 & $\mu$m \\
\hline
\end{tabularx}
\end{table}

\subsubsection{Multi-track simulation}
A major challenge in simulating multiple tracks in LPBF is the substantial number of time steps needed. To circumvent the cubic growth associated with the increasing number of temporal grid points and modes for moving source problems, we can leverage multiple S-T slabs to break down the original problem with a large number of time steps into smaller slabs. Consequently, this method keeps the total number of modes required in each slab beneath a reasonable threshold, thereby optimizing computational efficiency. The detailed algorithm of simulating multiple space-time (S-T) slabs for LPBF process is shown in \ref{app1}. Using this method, we first simulate a multi-track LPBF problem and analyze how the total number of slabs influences computational cost. The detailed setup can be found in Table \ref{tab:multiple slabs}. Note that we only simulate the printing process of the final layer in this section.

\begin{table}[!hbt]
\centering
\caption{Parameters used in the multi-track simulation}
\label{tab:multiple slabs}
\begin{tabularx}{\textwidth}{l|X|X|X}
\hline
\textbf{Parameter} & \textbf{Variable} & \textbf{Value} & \textbf{Units} \\
\hline
Laser power & \( P \) & 200 & W \\
Laser spot size radius & \( r \) & 50 & $\mu$m \\
Laser scan speed & \( V \) & 500 & mm s\(^{-1}\) \\
Absorptivity & \( \eta \) & 0.25 & 1 \\
Length & \( L \) & 1.5 & mm \\
Width & \( W \) & 1.5 & mm \\
Height & \( H \) & 1.5 & mm \\
Laser penetration depth & \( d \) & 50 & $\mu$m \\
Hatch space size & \( h_s \) & 50 & $\mu$m \\
Mesh size & \( h \) & 5 & $\mu$m \\
\hline
\end{tabularx}
\end{table}

\begin{figure}[!hbt]
\centering
\includegraphics[width=0.7\linewidth]{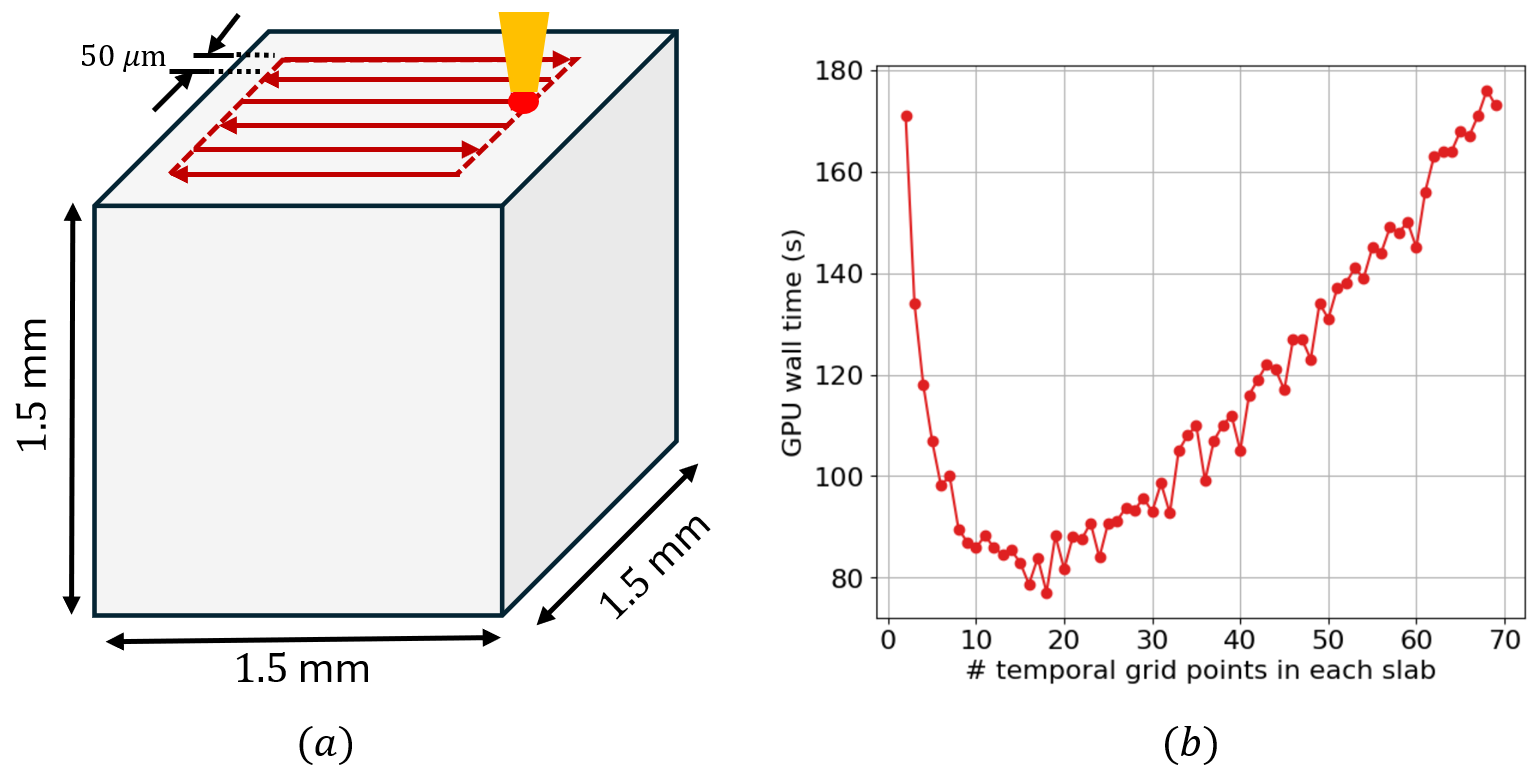}
\caption{(a) Multi-track simulation. (b) Influence of number of temporal grid points in each S-T slab on the computational cost.}
\label{multi_track}
\end{figure}

We use different numbers of temporal grids within each S-T slab and compare the computation cost, as shown in Fig. \ref{multi_track}. As can be seen from the figure, when each space-time slab contains around 20 temporal grid points, the computational efficiency is optimal.  Hence, choosing the optimal number of temporal elements inside each space-time slab is crucial for the overall performance of the TAPS solver for modeling LPBF process. We will adopt 20 temporal grid points per S-T slab as the default for the following multi-track LPBF simulations.

\begin{figure}[!hbt]
\centering
\includegraphics[width=1.0\linewidth]{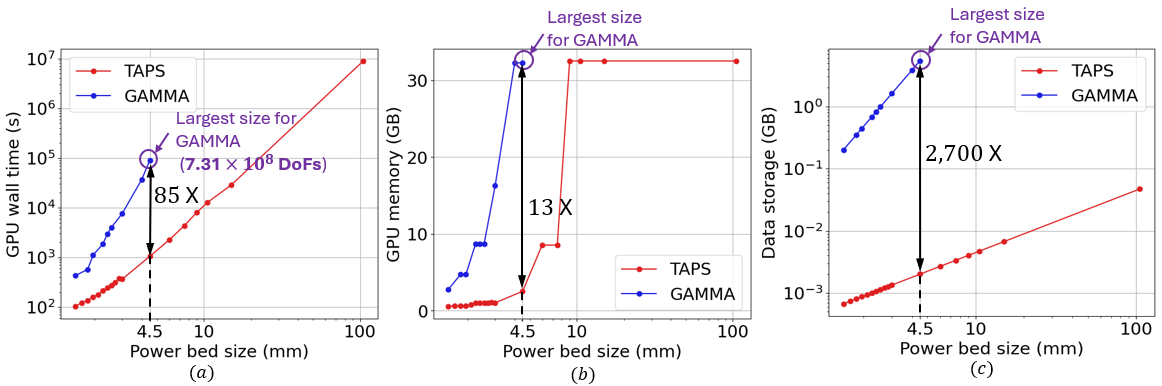}
\caption{Performance comparison of TAPS and GAMMA for powder bed with different sizes in terms of (a) computational time (b) GPU memory requirement (c) Data storage requirement for each time increment}
\label{gamma}
\end{figure}

Next, we compare the performance of TAPS versus the classical explicit finite difference method. To this aim, we use a GPU-accelerated and optimized finite difference code, GAMMA, to model the LPBF process \cite{liao2023efficient}. In this example, we increase the size of the domain while maintaining all other process parameters, as shown in Table \ref{tab:multiple slabs}. The corresponding computation time, GPU memory usage, and required data storage space are plotted in Fig. \ref{gamma}. 

Fig. \ref{gamma}(a) highlights the significant speed advantage of TAPS over GAMMA, especially as the size of the simulation domain increases. GAMMA only can simulate powder bed size up to $4.5^3$ mm$^3$ since the GPU memory can only handle up to $7.31 \times 10^8$ spatial DoFs. For the $4.5^3$ mm$^3$ case, TAPS is 85 times faster than GAMMA. On the other hand, TAPS is able to model $100^3$ mm$^3$ powder bed, with its speed benefits becoming even more evident for larger domains. Fig. \ref{gamma}(b) compares the memory requirements, where GAMMA experiences fast growth due to the cubic scaling of total spatial DoFs. In contrast, TAPS benefits from TD, requiring significantly less memory. TAPS uses 13 times smaller GPU memory compared to GAMMA for the $4.5^3$ mm$^3$ case. Additionally, TAPS can efficiently manage GPU memory usage for larger powder bed simulations by adopting different numbers of temporal grids in each S-T slab. Finally, Fig. \ref{gamma}(c) compares data storage needs where GAMMA's storage requirements grow cubically, whereas TAPS maintains a linear growth pattern. For the $4.5^3$ mm$^3$ case, the data storage of GAMMA is 2,700 times larger than TAPS.

\subsubsection{Large-scale multi-layer multi-track LPBF simulation}
\begin{figure}[!hbt]
\centering
\includegraphics[width=0.7\linewidth]{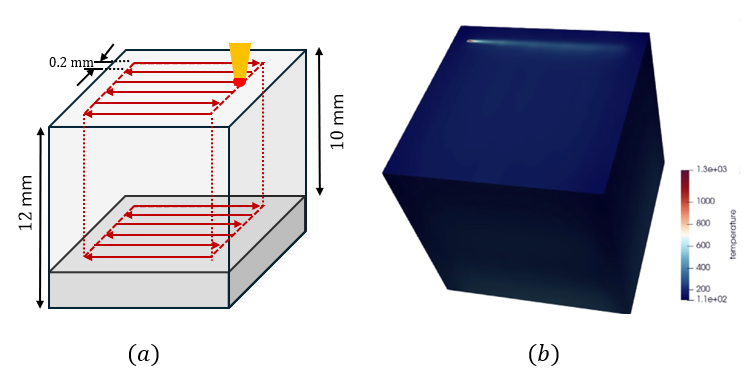}
\caption{(a) Problem statement: LPBF simulation. (b) Temperature solution for printing the final layer}
\label{lpbf}
\end{figure}

In this section, the proposed method is used to simulate a large-scale multi-layer multi-track LPBF process.  Element birth is used to model newly added layers in the process. Details on element birth can be found in \ref{app2}. As shown in Fig. \ref{lpbf} (a), the run scenario is the production of a 10 mm cube within a 12 mm powder bed domain. The base plate height is 2 mm. The tool path follows the pattern shown on the top surface. Material parameters are taken from Ti-6Al-4V \cite{leonor2024go}. The detailed parameters for the simulation setup are shown in Table \ref{tab:lpbf}.

\begin{table}[!hbt]
\centering
\caption{Parameters used in the large-scale LPBF simulation}
\label{tab:lpbf}
\begin{tabularx}{\textwidth}{l|X|X|X}
\hline
\textbf{Parameter} & \textbf{Variable} & \textbf{Value} & \textbf{Units} \\
\hline
Laser power & \( P \) & 200 & W \\
Laser spot size radius & \( r \) & 100 & $\mu$m \\
Laser scan speed & \( V \) & 500 & mm s\(^{-1}\) \\
Absorptivity & \( \eta \) & 0.25 & 1 \\
Laser penetration depth & \( d \) & 50 & $\mu$m \\
Layer thickness & \( h_l \) & 50 & $\mu$m \\
Hatch space size & \( h_s \) & 200 & $\mu$m \\
\hline
\end{tabularx}
\end{table}

To showcase the capabilities of our approach using TAPS, we employ a fine spatial mesh for the simulation. The spatial element size is $10\times10\times5 \mu m^3$. In classical numerical algorithms, this corresponds to $3.46\times10^9$ spatial DoFs, which is unmanageable for typical workstations due to the prohibitive RAM requirements.

The simulation result is shown in Fig. \ref{lpbf} (b), where the temperature of the last layer is plotted. In total, it costs 60.7 hrs to run the simulation. The maximum GPU memory usage is 8.11 GB. The final solution vector size is 1.35 GB. As a comparison, it's estimated that GAMMA will solve the same problem with the same spatial resolution in 3,485 days, with at least 120 GB GPU memory usage and 1.26 TB storage space to store the solution \cite{leonor2024go}. Consequently, TAPS achieves around 1,370 X speedup, 14.8 X memory footprint savings, and 955 X storage gain compared to the finite difference method.

\section{Discussion}
In the previous sections, we have shown that TAPS tackles two drawbacks of data-driven surrogate modeling approaches which use offline data generated through direct numerical simulation (DNS). Firstly, the proposed TAPS is data-free, which means that it does not require any training data. This is of crucial importance for applications that require ultra high-resolution simulations because offline training data generation can be extremely costly. Our method circumvents expensive offline DNS data generation by directly solving the governing equation. Secondly, TAPS enables solving ultra large-scale problems with significant speedup, minimal memory requirement, and substantial storage gain as compared to standard DNS techniques.

The computational speed of the current method can be further improved with the state-of-the-art high-performance numerical solvers and parallel computing on multiple GPUs. Right now, the TAPS linear systems of equations are solved on CPUs which results in additional overhead. With more sparse direct solvers/iterative schemes becoming available on GPU, we expect a further speedup of the current program. Moreover, parallel computing using multiple GPUs can be achieved using Message Passing Interface (MPI) \cite{jacobsen2010mpi}. For ultra large-scale analysis where each dimension contains millions of nodes, an efficient iterative solver with a suitable preconditioner needs to be developed.

Variational multiscale methods can be used to further extend the capabilities of the current method to tackle zetta-scale space-time problems \cite{hughes1998variational,leonor2024go}. Moreover, one major computational cost for the current method originates from the increased number of decomposition modes for a large number of time steps. This can be avoided by leveraging coordinate transformation techniques where the moving source can be transformed into a fixed one. As a result, we expect to greatly improve the computational performance of the current method. Irregular geometry can also be considered using immersed finite element techniques or the Solid Isotropic Material with Penalization (SIMP) method in topology optimization \cite{li2023convolution, wang2013modified,liu2006immersed,liu2007mathematical,kopacz2012nanoscale}.

\section{Conclusion}

In this paper, we propose TAPS as a data-free predictive scientific AI model to  simulate  ultra large-scale physical problems. This method eliminates the traditional necessity for offline training data generation, thereby exhibiting substantial speedup, memory efficiency, and storage gain as opposed to data-driven methods, making previously unsolvable large-scale and high-dimensional problems manageable. The convergence of the TAPS solver is numerically investigated. As a demonstration of the capabilities of TAPS, we showcase the application of the TAPS solver for a multi-layer multi-track additive manufacturing problem that is intractable with classical numerical algorithms. 

TAPS is well suited for a broad range of science or engineering problems where: 1) the finite element method and other
conventional numerical methods are unsuitable due to excessively long simulation times or high RAM and storage demands
needed to achieve high accuracy, 2) the model must accommodate design parameters as inputs, or 3) fast prediction is
required once the model is obtained. The INN hierarchical neural network interpolants, particularly C-HiDeNN used by TAPS, demonstrate superior performance compared to other machine learning models. For the solving tasks, it has shown superior performance compared to physics-informed neural network (PINN) \cite{raissi2019physics}, CP-PINN \cite{vemuri2025functional}, and Kolmogorov–Arnold Networks (KAN) \cite{liu2024kan} with orders of magnitude faster solution time, higher accuracy, and better scalability to ultra large-scale and high-dimensional PDEs \cite{guo2025interpolation}.  INN interpolants can also be effectively used in data-driven training tasks and show better training accuracy compared to MLP, SIREN \cite{sitzmann2020implicit} and KAN \cite{park2024engineering, guo2025interpolation}.

\begin{figure}[!hbt]
\centering
\includegraphics[width=0.7\linewidth]{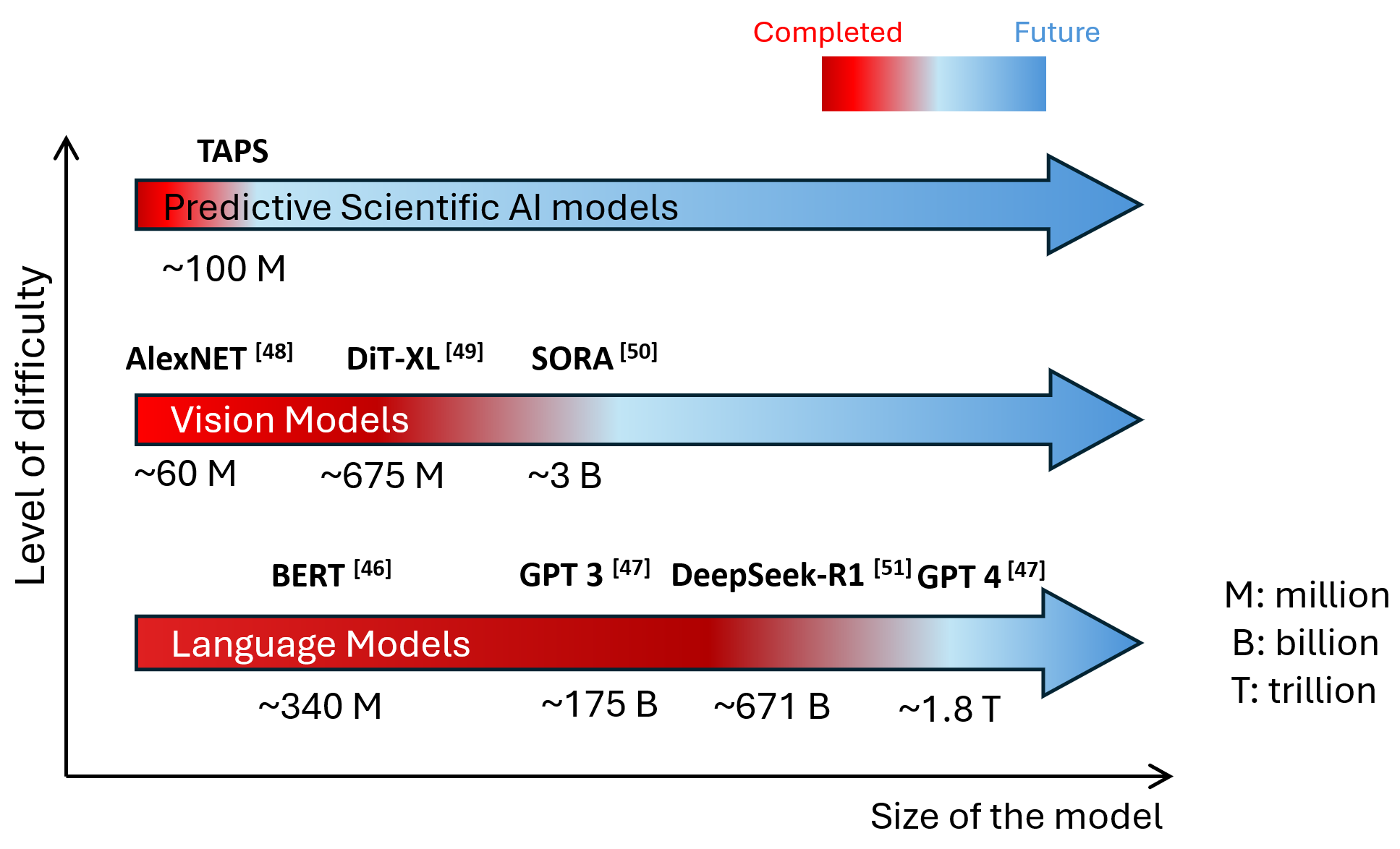}
\caption{Evolution of AI models for different tasks}
\label{evolution}
\end{figure}

As illustrated in Fig. \ref{evolution}, the significance of this work in the area of predictive scientific AI models aligns with the trend in other areas in AI, such as language and vision AI models. The evolution of language models has seen dramatic growth, beginning with foundational models like BERT \cite{devlin2019bert},  followed by the GPT series \cite{radford2018improving}, which expanded transformer architecture to hundreds of billions of parameters, showcasing powerful generative capabilities. In vision models, AlexNet \cite{krizhevsky2012imagenet} marked a breakthrough, while advancements like DIT-XL \cite{peebles2023scalable} and SORA \cite{liu2024sora} integrated diffusion models to handle more complex and challenging visual tasks. This trajectory of increasing scale and sophistication from its network architecture (i.e., transformer of language models and diffusion of vision models) is mirrored in predictive scientific AI where TAPS represents a significant advancement in its network architecture, INN.

A major critical issue in the emerging large AI models is a more sophisticated model will generally lead to a larger amount of training data, more expensive training costs, and longer inference time. The advent of DeepSeek R1 breaks this rule since it has fewer parameters, much less training cost, faster inference speed, yet still comparable accuracy compared to other state-of-the-art models due to its novel architecture and training techniques such as distillation methods \cite{guo2025deepseek}.  For predictive scientific AI, we face even more pronounced challenges due to strict accuracy demands and the necessity for high-resolution physics for large-scale problems. As a result, the future of predictive scientific AI is still largely untapped. TAPS provides a promising solution to these emerging challenges by delivering a highly accurate, exceptionally fast, and memory and storage efficient scientific AI model.

In conclusion, the proposed TAPS computational framework  offers substantial enhancements in computational efficiency, memory consumption, and storage demands for science and engineering simulations.  As a result, TAPS paves a new path to address future challenges in ultra large-scale simulations pertinent to complex predictive scientific AI models.

\appendix

\section{Solving nonlinear S-P-T PDEs: solution dependent material properties}
\label{app:_nonlinear_PDE}

The algorithm \ref{alg:linear_td} works for linear PDEs where the PDE coefficients remain constant. However, in many engineering applications, the PDE coefficients can be solution-dependent. For instance, material properties such as heat conductivity and heat capacity can be a function of temperature in additive manufacturing. In these cases, the PDE becomes non-linear which requires an efficient solution scheme. In this section, we solely focus on the space-time problems formulation of a nonlinear PDE. As a result, the product of density and heat capacity  $\rho c_p (u)$  and conductivity $k(u)$ are no longer temperature independent as in Eq. \ref{pde_eqn}. Similar to the linear problem shown before, the generalized Galerkin weak form is used to solve this equation.

\begin{equation}
    \int_\Omega\delta u\nabla_{x_t}\left[\rho c_p(u)u\right]d\Omega-\int_\Omega{\nabla_{\bm{x}_s}\delta u}\cdot k(u)\nabla_{\bm{x}_s} ud\Omega+\int_{\partial \Omega_{\bm{x}_s}\otimes\Omega_t}\delta u\bm{q}\cdot\boldsymbol{n}d\bm{s}d\Omega_{x_t}=\int_\Omega{\delta u}bd\Omega
\label{nonlinearG}
\end{equation}
where $\bm{q}$ is the heat flux on the Neumann boundary. Since Eq. \ref{nonlinearG} is a space-time integral, classical time-stepping based methods can’t be directly used to update material parameters. Here we propose a global-local approach similar to the Large Time Increment (LATIN) method to effectively solve the above equations \cite{ladeveze2016reduced}. 

\begin{figure}[!hbt]
\centering
\includegraphics[width=0.3\linewidth]{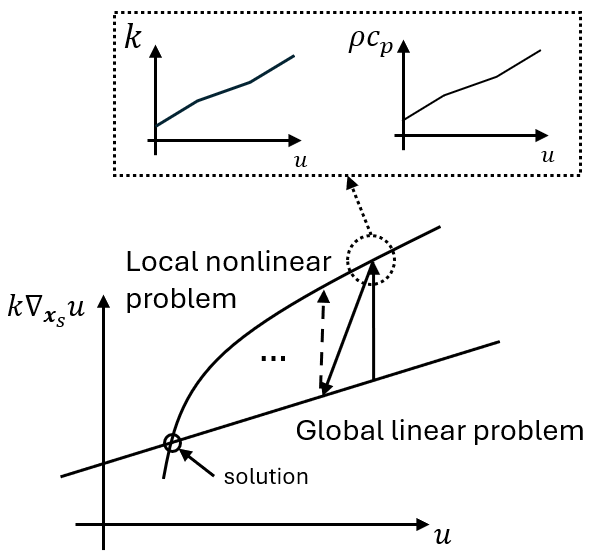}
\caption{Global-local approach for nonlinear TAPS solver}
\label{latin}
\end{figure}

As shown in Fig. \ref{latin}, we split the nonlinear problem into 2 stages, a linear global stage and a nonlinear local update stage. In the global stage, we assume the spatio-temporal $k(\bm{x}_s, x_t)$ and $\rho c_p (\bm{x}_s, x_t)$ are known. As a result, we treat the global problem as a linear problem and obtain $u(\bm{x}_s, x_t)$ using the previously proposed method for linear problems. 
After $u(\bm{x}_s, x_t)$ is updated in the global stage, we update $k(u)$  and $\rho c_p (u)$ locally at each Gauss integration point according to material models $k(u)$ and $\rho c_p (u)$. We repeat the global-local iteration until the variation of $k(u)$  and $\rho c_p (u)$ between consecutive iterations meets the convergence criteria. The algorithm is summarized in Algorithm \ref{alg:nonlinear_td}:

\begin{algorithm}
\caption{Nonlinear TAPS solution scheme: PDE with solution dependent coefficients}
\label{alg:nonlinear_td}
\begin{algorithmic}[1]
\State Initialize solution matrices with random values and update $\rho_{c_p}(\bm{x}_s, x_t)$ and $k(\bm{x}_s, x_t)$.

\For{$iter_{\gamma} = 1$ to $iter_{\gamma_{max}}$}
    \For{$iter = 1$ to $iter_{max}$} \Comment{Global linear problem}
        \State Update $\rho_{c_p}(\bm{x}_s, x_t)$ and $k(\bm{x}_s, x_t)$
        \State Use Algorithm 1 to solve solution $u(\bm{x}_s, x_t)$
        
        \For{$i = 1$ to integration points} \Comment{Local nonlinear problem}
            \State $\rho_{c_p}(\bm{x}_s, x_t) = \rho_{c_p}[u(\bm{x}_s, x_t)]$
            \State $k(\bm{x}_s, x_t) = k[u(\bm{x}_s, x_t)]$
        \EndFor
        
        \State Check convergence
    \EndFor
\EndFor
\end{algorithmic}
\end{algorithm}

\section{Mode compression}
\label{app1}
One significant challenge in multi-track simulation in LPBF is the huge number of time steps required. It is impossible to resolve all the time steps with only a single space-time (S-T) slab. Hence, we split the whole layer scan into multiple S-T slabs and relate each S-T slab using the following equation.
\begin{equation}
    ^{[\mathcal{T}+1]}u(\bm{x_s},x_t)={}^{[\mathcal{T}]}{u}(\bm{x}_s, x_t^{max})+^{[\mathcal{T}+1]}u_0(\bm{x_s},x_t)
\label{app1_eq1}
\end{equation}
where $^{[\mathcal{T}+1]}u(\bm{x_s},x_t)$ refers to the solution at $(\mathcal{T}+1)$-th space-time slab; $^{[\mathcal{T}+1]}u_0(\bm{x_s},x_t)$ refers to the solution of the homogeneous initial value problem of $(\mathcal{T}+1)$-th space-time slab; ${}^{[\mathcal{T}]}{u}(\bm{x}_s, x_t^{max})$  is the solution of $\mathcal{T}$-th space-time slab at the last time increment. As can be seen from Eq. \ref{app1_eq1}, we impose the last time increment solution of previous space-time slab as the initial condition for the next space-time slab. This is efficiently implemented by adding the TD form of the last increment as new modes in the current space-slab solution. However, for large-scale computations requiring thousands of slabs, directly concatenating modes can result in substantial storage demands.

In mode compression, we aim to compress the number of modes for ${}^{[\mathcal{T}]}{u}(\bm{x}_s, x_t^{max})$ because of its spatial dependence and naturally low-dimensional structure. Consequently, it can be effectively decomposed using only a few modes. Denote the TD form of the last time step solution of the previous space-time slab as ${}^{[\mathcal{T}]}{u}(\bm{x}_s, x_t^{max})^{TD}$, we aim to find a compact form that can be represented with much fewer number of modes  ${}^{[\mathcal{T}]}{u}(\bm{x}_s, x_t^{max})^{TD}_F$. For notation simplicity, we omit $x_t^{max}$ in the following equations. Consequently, the mode compression problem can be written as:
\begin{equation}
    ^{[\mathcal{T}]}u(\bm{x_s})_F^{TD}-^{[\mathcal{T}]}u(\bm{x_s})^{TD}=0
\end{equation}
The weighted sum residual form is used to approximate $^{[\mathcal{T}]}u(\bm{x_s},x_{t_{-1}})^{TD}_{F}$:
\begin{equation}
\int_{\Omega_{\bm{x}}}\delta^{[\mathcal{T}]}u(\bm{x_s})_F^{TD}\cdot\left[^{[\mathcal{T}]}u(\bm{x_s})_F^{TD}-^{[\mathcal{T}]}u(\bm{x_s})^{TD}\right] d \bm{x}_s= 0
\label{app1_eq3}
\end{equation}
Eq. \ref{app1_eq3} can be efficiently solved using Algorithm \ref{alg:linear_td}.

\section{Element birth}
\label{app2}
In the LPBF process, once the printing is finished for the current layer, a new layer of powder is deposited on top of the existing layer. This necessitates modeling the new layer with additional elements. Various studies have investigated different approaches for element birth techniques. While some researchers opt for activating small sections of geometry incrementally, others apply the technique by spreading the deposition across an entire layer or multiple layers simultaneously. The most widely adopted approach is to activate an entire layer and then scan the heat source over it  \cite{williams2018pragmatic}.

In TAPS, we propose a new scheme to generate new layers of elements. In this scheme, new elements are added only in the $x_3$ direction, since the plan dimension doesn’t change in the printing process. Therefore, as opposed to full-scale classical numerical methods, TAPS enables marginal overhead in generating new layers of elements with extra grid points added only in the $x_3$ dimension. The solution scheme for multi-layer multi-track LPBF simulation using TAPS can be summarized in Algorithm \ref{alg:multi_layer_lpbf}.

\begin{algorithm}
\caption{Multi-layer multi-track LPBF simulation using TAPS}
\label{alg:multi_layer_lpbf}
\begin{algorithmic}[1]
\For{$n_{layer} = 1$ to $n_{layerTotal}$}
    \State Initialize solution matrices with random values for the new layer
    \State Compute the updated stiffness matrix and force vector for the $x_3$ direction
    \For{$n_{track} = 1$ to $n_{tracktotal}$}
        \For{$iter = 1$ to $iter_{max}$} 
            \For{$d = 1$ to dimension}
                \State Compute solution vectors according to Algorithm 1 or 2
            \EndFor
            \State Check convergence
        \EndFor
        \State Compress modes
        \State Concatenate compressed modes to previous tracks as new modes
    \EndFor
    \State Compress modes
    \State Concatenate compressed modes to previous layers as new modes
    
\EndFor
\end{algorithmic}
\end{algorithm}






\bibliography{reference}

\end{document}